\documentclass[letter,longauth]{aa} 

\usepackage{style_paper}

\usepackage{graphicx}
\usepackage{txfonts}
\begin{document}

   \title{A kinematically detected planet candidate in a transition disk \thanks{The ALMA 12CO molecular line emission cube is only available in electronic form at the CDS via anonymous ftp to \url{cdsarc.cds.unistra.fr} (130.79.128.5) or via \url{https://cdsarc.cds.unistra.fr/cgi-bin/qcat?J/A+A/}}}
   \author{J. Stadler\inst{1,2} , M. Benisty\inst{1,2}, A. Izquierdo\inst{3,4}, S. Facchini\inst{5}, R. Teague\inst{6}, N. Kurtovic\inst{7}, P. Pinilla\inst{8}, J. Bae\inst{9},  \\
   M. Ansdell\inst{10}, R. Loomis\inst{11}, S. Mayama\inst{12}, L. M. Perez \inst{13,14}, L. Testi\inst{3}}

   \institute{
   $^{1}$ Laboratoire Lagrange, Université Côte d’Azur, CNRS, Observatoire de la Côte d’Azur, 06304 Nice, France;\\ \email{jochen.stadler@oca.eu} \\
   $^{2}$ Univ. Grenoble Alpes, CNRS, IPAG, 38000 Grenoble, France\\
   $^{3}$ European Southern Observatory, Karl-Schwarzschild-Str. 2, D-85748 Garching bei München, Germany\\
   $^{4}$ Leiden Observatory, Leiden University, P.O. Box 9513, NL-2300 RA Leiden, The Netherlands\\
   $^{5}$ Universit\'a degli Studi di Milano, via Celoria 16, 20133 Milano, Italy \\
   $^{6}$ Department of Earth, Atmospheric, and Planetary Sciences, Massachusetts Institute of Technology, Cambridge, MA 02139, USA\\
   $^{7}$ Max Planck Institute for Astronomy, K\"onigstuhl 17, 69117, Heidelberg, Germany\\
   $^{8}$ Mullard Space Science Laboratory, University College London, Holmbury St Mary, Dorking, Surrey RH5 6NT, UK\\
   $^{9}$ Department of Astronomy, University of Florida, Gainesville, FL 32611, USA \\
   $^{10}$ NASA Headquarters, 300 E Street SW, Washington, DC 20546, USA\\
   $^{11}$ National Radio Astronomy Observatory, Charlottesville, VA 22903, USA \\
   $^{12}$ The Graduate University for Advanced Studies, SOKENDAI, Shonan Village, Hayama, Kanagawa 240-0193, Japan\\
   $^{13}$ Departamento de Astronom\'ia, Universidad de Chile, Camino El Observatorio 1515, Las Condes, Santiago, Chile\\
   $^{14}$ N\'ucleo Milenio de Formaci\'on Planetaria (NPF), Avenida España 1680, Valparaíso, Chile \\ }

   \date{Received 7 November 2022 / Accepted 2 January 2023}


  \abstract
   {Transition disks are protoplanetary disks with inner cavities possibly cleared by massive companions. Observing them at high resolution is ideal for mapping their velocity structure and probing companion--disk interactions.}
   {We present Atacama Large Millimeter/submillimeter Array (ALMA) Band 6 dust and gas observations of the transition disk around RXJ1604.3–2130\,A, known to feature nearly symmetric shadows in scattered light, and aim to search for non-Keplerian features.}
  {We studied the $^{12}$CO line channel maps and moment maps of the line-of-sight velocity and peak intensity. We fitted a Keplerian model of the channel-by-channel emission to study line profile differences and produced deprojected radial profiles for all velocity components.}
   {The $^{12}$CO emission is detected out to $R\sim$1.8\arcsec{} (265\,au). It shows a cavity inward of 0.39\arcsec{} (56\,au) and within the  dust continuum ring (at $\sim$0.56\arcsec{}, i.e., 81\,au).
   Azimuthal brightness variations in the $^{12}$CO line and dust continuum are broadly aligned with the shadows detected in scattered-light observations. We find a strong localized non-Keplerian feature toward the west within the continuum ring (at ${R=41\pm10}$ au and ${PA=280\pm2^\circ}$). It accounts for $\Delta v_\phi/v_\mathrm{kep}\sim0.4$ or $\Delta v_z/v_\mathrm{kep}\sim0.04$, depending on if the perturbation is in the rotational or vertical direction.
   A tightly wound spiral is also detected and extends over $300^\circ$ in azimuth, possibly connected to the localized non-Keplerian  feature. Finally, a bending of the iso-velocity contours within the gas cavity indicates a highly perturbed inner region, possibly related to the presence of a misaligned inner disk. }
   {While broadly aligned with the scattered-light shadows, the localized non-Keplerian feature cannot be solely due to changes in temperature. Instead, we interpret the kinematical feature as tracing a massive companion located at the edge of the dust continuum ring. We speculate that the spiral is caused by buoyancy resonances driven by planet--disk interactions.
   However, this potential planet at $\sim$41\,au cannot explain the gas-depleted cavity, the low accretion rate, and the misaligned inner disk, which suggests the presence of another companion closer in.
   }


   \titlerunning{A kinematically detected planet candidate in J1604}
   \authorrunning{Stadler et al.}
   \maketitle
%
\section{Introduction}
Planet formation appears to be a robust and efficient process, occurring around both single and multiple stellar systems \citep{Kostov2016} in protoplanetary disks.
The advent of high resolution imaging facilities has demonstrated that nearly all bright and extended disks show substructures, in particular in the small (micron-sized) and large (millimeter-sized) dust tracers seen through scattered and thermal light, respectively \citep[e.g.,][]{Andrews_ea_2018,Long_ea_2018,Rich2022,PPVII_Benisty,Bae_PPVII}. Such high resolution studies applied to gas tracers allow the overall physical conditions in the disk to be probed, such as its temperature structure, its surface height \citep{Rich2021,Law_ea_2021b}, and pressure variations \citep{Teague_ea_2018b,Teague_ea_2018c,Rosotti_ea_2020}. Studies of the disk density and the velocity structure reveal a great complexity, including localized non-Keplerian features that can be attributed to embedded massive protoplanets \citep[e.g.,][]{PPVII_Pinte, Wolfer_ea_2022}. Such perturbations from smooth density and velocity distributions can directly constrain planet formation since they are expected to leave clear signatures on the disk structure \citep[e.g.,][]{Perez_ea_2015,Yun_ea_2019}. For example, the mapping of spiral wakes \citep{Calcino2022} and the detections of so-called Doppler flips \citep[change of sign in the non-Keplerian feature; e.g.,][]{Casassus_Perez_2019,Norfolk2022}, meridional flows within dust-depleted gaps \citep{Teague_ea_2019b}, and a velocity perturbation associated with a circumplanetary disk candidate \citep{Bae_ea_2022} enable us to take a closer look at planet--disk interaction processes. While most localized kinematical perturbations are analyzed empirically, statistical methods for quantifying their significance have been developed and led to the detection of localized signatures possibly associated with unseen planets \citep{Izquierdo_ea_2021,Izquierdo_ea_2022}. Prime targets to search for protoplanets still embedded in their birth environment are the so-called transition disks. As in PDS70 \citep{Keppler_ea_2019} or AB\,Aur \citep{Tang2017}, these disks host a dust-depleted cavity that has possibly been cleared by massive companions \citep{Zhu2011}.

\begin{figure*}[t]
\centering
\includegraphics[width=0.9\linewidth]{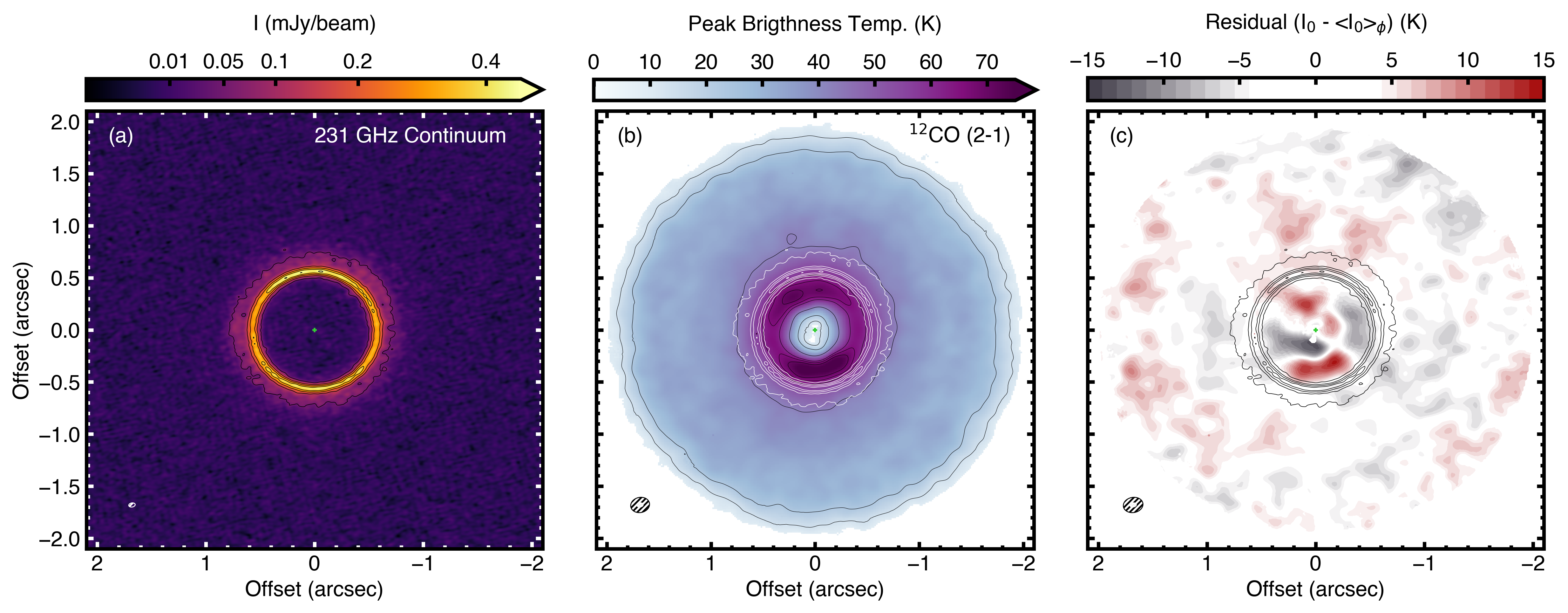}
\caption{ALMA observations of J1604. Panel\,\textbf{(a)}: 231 GHz dust continuum.\ The solid black contours are drawn at [5, 15, 25, 35, 45]$\sigma$, and the image is plotted with a power-law scaling of $\gamma=0.6$. Panel \textbf{(b)}: $^{12}$CO peak brightness temperature map computed from $I_\mathrm{0}$ using the Planck law with black solid contours drawn at [5, 10, 20, 40, 60, 65, 70]\,$\sigma$. Pixels below $5\sigma$ are masked. Panel \textbf{(c)}: Peak intensity residuals after subtracting an azimuthally averaged radial profile from the data. We have adjusted the color scale such that residuals smaller than $1\sigma$ are white. The beam sizes are shown in the lower-left corner, and the position of the star is marked by a green cross. In panels \textbf{(b)} and \textbf{(c)}, we have overlaid the continuum contours in white and black, respectively.}
\label{fig:intensities}
\end{figure*}

In this Letter we focus on RXJ1604.3-2130\,A \citep[d=144.6\,pc, $1.24\,M_\odot$;][respectively]{Gaia_dr3_2022,Manara_ea_2020}, hereafter J1604, one of the brightest protoplanetary disks of the Upper Scorpius Association in the millimeter regime \citep{Barenfeld2016}, which exhibits a prominent cavity in the dust continuum and CO line emission \citep{Zhang_ea_2014,Dong2017,vanderMarel_ea_2021a}. J1604 has a stellar companion located at $\sim$2300\,au, itself a binary with separation 13\,au \citep{Koehler_ea_2000}. The outer disk of J1604 was resolved with the Atacama Large Millimeter/submillimeter Array \citep[ALMA;][]{Mayama2018} and the Spectro-Polarimetric High-contrast Exoplanet REsearch (SPHERE) instrument on the Very Large Telescope \citep[VLT;][]{Pinilla2015}, indicating a nearly face-on geometry. Complementary observations are indicative of a misaligned inner disk with respect to the outer disk. Its variable light curve is that of an irregular dipper \citep{Ansdell2020}, infrared scattered-light observations show the presence of two shadows with variable morphology on timescales possibly shorter than a day \citep{Pinilla2018}, and ALMA $^{12}$CO (J=3--2) line observations show deviations from Keplerian rotation in the cavity \citep{Mayama2018}. The position of the scattered-light shadows are suggestive of a large misalignment ($\sim$70-90$^\circ$). Measurements of the projected rotational velocity ($v$\,sin$i$) indicate that the star is aligned with the inner disk, and thus misaligned with the outer disk \citep{Sicilia2020}.

In this work, we present new ALMA observations of J1604 and focus on the kinematics of the $^{12}$CO (J=2--1) line. Section\,\ref{sec:calib} presents the observations and Sects.\,\ref{sec:method} and \ref{sec:results} our methodology and results, respectively. Section\,\ref{sec:discussion} provides a discussion of the results and Sect.\,\ref{sec:conclusion}, our conclusions.

\section{Observations, calibration, and imaging}
\label{sec:calib}
We present new ALMA Band~6 observations (2018.1.01255.S; PI:~Benisty) with five executions spread over two years, obtained on 2019 April 4, 2019 July 30 and 31, 2021 April 29, and 2021 September 27. The spectral setup was designed for continuum detection but includes the $^{12}$CO J=2-1 line.
The data were combined with archival data from program 2015.1.00964.S (PI Oberg; see Tab.\,\ref{tab:obsdata}).
The data calibration and imaging were performed following the procedure of \cite{Andrews_ea_2018}, with \texttt{CASA v.5.6.1} \citep{McMullin2007}, and is detailed in Appendix \ref{app:obs}. The synthesized beam of the $^{12}$CO line and dust continuum images is 0.18\arcsec{}\,x\,0.15\arcsec{} (102$^\circ$) and 0.060\arcsec{}\,x\,0.039\arcsec{} (- 78$^\circ$), respectively. The rms in a line-free channel was measured to be 1.1\,mJy\,beam$^\mathrm{-1}$ (4.3\,K) for CO and 10\,µJy\,beam$^\mathrm{-1}$ for the dust continuum. Figure\,\ref{fig:intensities} shows the dust continuum map (left), which displays a cavity and a bright dust ring peaking at $R\sim$0.56\arcsec{} ($\sim$81\,au), and the $^{12}$CO peak brightness temperature map (center), which indicates a smaller cavity in gas, with a peak at $R\sim$0.39\arcsec{} ($\sim$56\,au). A selection of channel maps can be found in Fig.\,\ref{fig:app_channel_maps_compared}.

\section{Methodology}
\label{sec:method}
\paragraph{Channel maps model.}
\label{sec:discminer}

To model the disk line intensity and kinematics, we used the \texttt{discminer} package of \cite{Izquierdo_ea_2021}. The code uses parametric prescriptions for the line peak intensity, line width, rotational velocity, and disk emission height to produce channel maps; it uses \texttt{emcee} \citep{emcee2013} to maximize a $\chi^2$ log-likelihood function of the difference between the model and input intensity for each pixel in a channel map.
To prescribe the model intensity, we used a generalized bell kernel function of the disk cylindrical coordinates (R, z):
\begin{equation}
    I_\mathrm{m} (R, z; v_\mathrm{ch}) = I_\mathrm{p}(R, z)\left(1\,+\,\left|\frac{v_\mathrm{ch} - v}{L_\mathrm{w}(R, z)}\right|^{2L_\mathrm{s}}\right)^{-1},
\end{equation}
where $I_\mathrm{p}$ is the peak intensity, $L_\mathrm{w}$ is half the line width at half power, hereafter ``the line width,'' and $L_\mathrm{s}$ is the line slope. The $v_\mathrm{ch}$ is the channel velocity at which the intensity is computed and $v$ the observed Keplerian line-of-sight velocity. As the disk is nearly face-on, the code is unable to infer an emission height, and we therefore assumed a flat emission surface. We additionally fixed the inclination, $i$, of the disk to the one inferred from the dust continuum \citep[$i=6.0^\circ$;][]{Dong2017} to break the degeneracy of $M_\star \cdot \sin{i}$. The fitting procedure and the Markov chain Monte Carlo (MCMC) search are explained in detail in Appendix \ref{app:mcmc_model}, where the functional form of each model parameter together with its best-fit parameters are summarized in Table \ref{table:best_fit_param}. We compare selected channel maps to the best-fit model using these parameters in Fig.\,\ref{fig:app_channel_maps_compared}.

\begin{figure*}[t]
\centering
\includegraphics[width=0.9\linewidth]{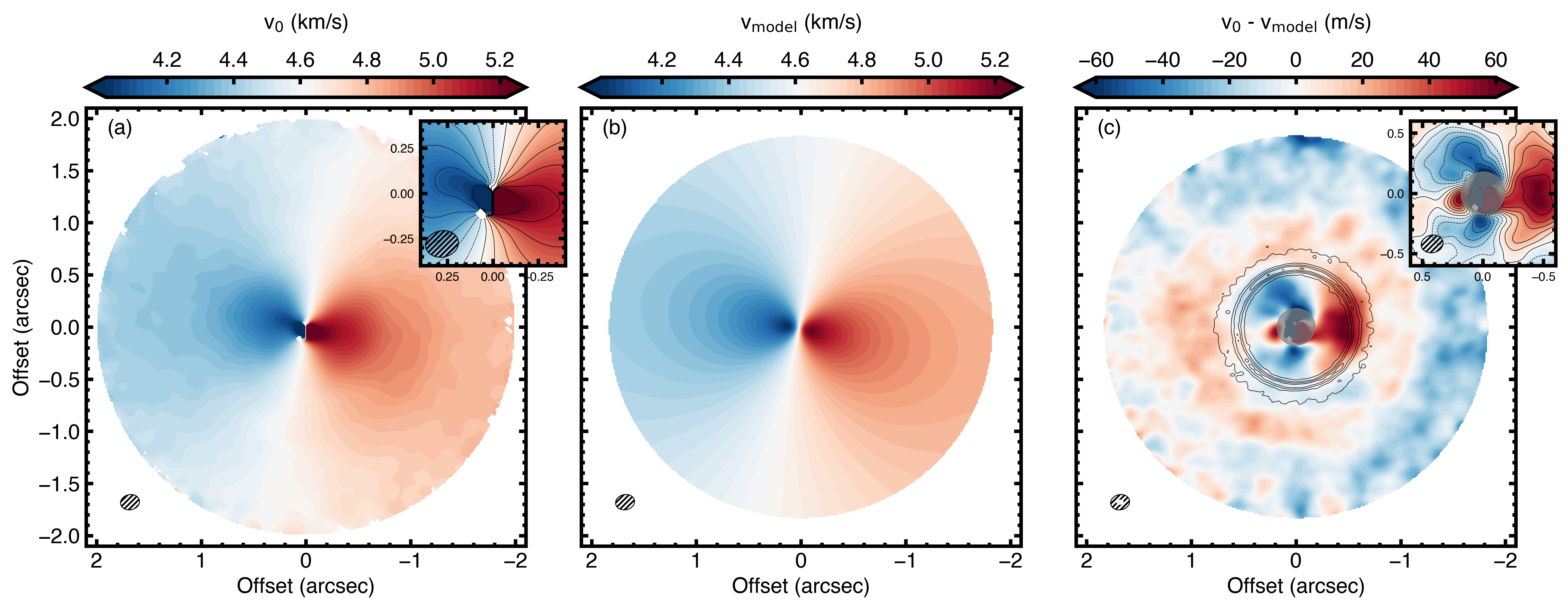}
\caption{Line-of-sight velocity maps for $v_\mathrm{0}$ data \textbf{(a)} and the \texttt{discminer} model, $v_\mathrm{mod}$ \textbf{(b)}. Panel \textbf{(c)}: Velocity residual map after subtracting $v_\mathrm{mod}$ from $v_\mathrm{0}$. The dust continuum is overlaid in solid contours with equal levels, as in Fig. \ref{fig:intensities}. The innermost region was masked during the fit by one beam size in radius, shown as the gray shaded ellipse. The insets in panels \textbf{(a)} and \textbf{(c)} zoom into the innermost region of the disk to highlight the non-Keplerian velocities. Contours are drawn at v$_\mathrm{sys}=(4.62\,$±\,0.60)\,\kmsec in steps of 0.1\,\kmsec and from -60 to 60\,\msec in steps of 10\,\msec, respectively. All maps show the synthesized beam for CO (black) and the continuum (white) in the lower-left corner.  The beams are masked where the CO peak intensity falls below the 5$\sigma$ level for panel \textbf{(a) }and where $R>R_\mathrm{out}$ for the rest.}
\label{fig:centroid}
\end{figure*}

\paragraph{Moment maps.}
\label{sec:moments}
The moment maps were computed with \texttt{bettermoments} \citep{bettermoments}. Since the $^{12}$CO line emission is optically thick, we fitted the following line profile to both the data and model channel maps:
\begin{equation}
\label{eq:Gauss_thick}
    I (v) = I_\mathrm{0}\cdot \frac{1 - \exp\left(-\tau\left(v\right)\right)}{1 - \exp(-\tau_\mathrm{0})}\;\,\text{with}\;\,\tau=\tau_0\exp{\left(\frac{-(v-v_0)^2}{\Delta V^2}\right)},
\end{equation}
where $I_\mathrm{0}$ is the peak intensity of the line and the optical depth, $\tau\,(v)$, varies like a Gaussian; $v_0$ is the line centroid, $\tau_0$ the peak optical depth, and $\Delta V$ the width of the line -- where the full width at half maximum (FWHM) = $2\sqrt{\mathrm{ln}2}\Delta V$--  as used in \cite{Teague_ea_2022b}. In Fig.\,\ref{fig:intensities}\,(b) we show $I_\mathrm{0}$ for $^{12}$CO in units of brightness temperature. The corresponding $v_\mathrm{0}$ maps for the data and model are displayed in panels  (a)\,and\,(b)  of Fig.\,\ref{fig:centroid},  respectively. The moment maps for $\Delta V$ and $\tau_0$, as well the error of the line centroid fitting, $\delta v_0$, can be found in Fig.~\ref{fig:app_moment_maps}.

\section{Results}
\label{sec:results}
\subsection{Dust and gas radial and azimuthal brightness profiles}
\label{sec:gas_emission}
Figure \ref{fig:intensities} shows the 1.3\,mm dust continuum together with the peak brightness temperature map, $I_0$, of the $^{12}$CO (J=2-1) line emission. Both dust and gas tracers show a cavity, and the $^{12}$CO (J=2-1) line emission extends inward of the dust continuum, as expected if the continuum ring results from dust trapping \citep[e.g.,][see our Fig.  \ref{fig:app_I_azim_avg}]{Facchini2018}. We note that the $^{12}$CO cavity appears asymmetric with respect to the position of the star and that we observe a gap-like feature in the $^{12}$CO peak intensity map at $R\sim$1.2\arcsec{}, apparent as a plateau of $I_0\approx31$\,mJy\,beam$^{-1}$ stretching over $\Delta R\approx$\,0.1\arcsec{} (Fig.\,\ref{fig:app_I_azim_avg}).
Interestingly, the disk shows significant azimuthal intensity variations (34\% at R=0.56\arcsec{}; 19\% at 0.39\arcsec{} for continuum and gas, respectively; see also Fig.~\ref{fig:app_azim_int}). Figure\,\ref{fig:intensities}\,(c) shows the residuals obtained after subtracting an azimuthally averaged radial profile from the $^{12}$CO peak brightness temperature map. Azimuthal variations are clearly apparent within the dust cavity, with residual values of $>10\sigma$. The fainter regions, distributed along the east-west direction, are broadly aligned with fainter regions seen in the continuum (see the contours of Figs.\,\ref{fig:intensities}(b) and \ref{fig:app_azim_int}) and with the shadows reported in scattered light \citep[][; Kurtovic et al. in prep]{Pinilla2018}.

\subsection{Kinematical features}
\label{sec:kinematics}
\subsubsection{Localized velocity residuals}
\label{sec:cent_resid}
\begin{figure}[t]
\centering
\includegraphics[width=\linewidth]{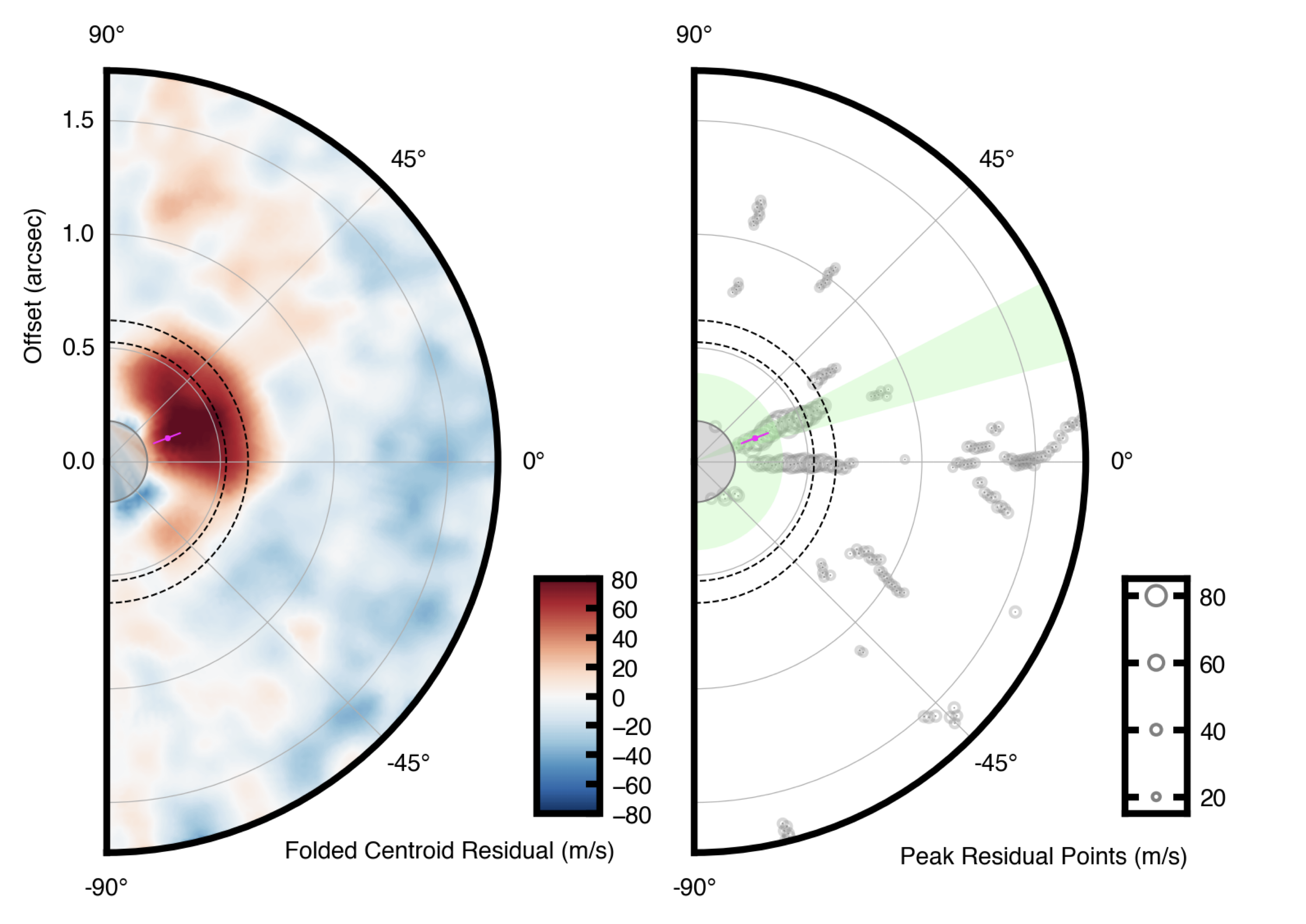}
\caption{Folded velocity residuals (left) and detected clusters of peak velocities (right) in the disk reference frame. The green shaded annulus and wedge in the right plot mark the significant clusters in radius and azimuth, respectively. The position of the localized velocity perturbation inferred from these clusters is marked with a magenta point with error bars. The gray region (one  beam size in radius) indicates the masked area, and the black dashed lines the FWHM of the dust ring.}
\label{fig:polar_plot_clusters}
\vspace{-2.5mm}
\end{figure}

The centroid residual map in Fig.\,\ref{fig:centroid}\,(c) shows a prominent, localized non-Keplerian velocity feature of $\delta v\approx$\,60\,\msec, between $\sim$0.35\arcsec{} and 0.55\arcsec{} (i.e., 50-80\,au), which is at the edge of the dust continuum ring and oriented at PA\,$\approx(270\pm15)^\circ$. To assess its significance, we followed the variance peak method from \cite{Izquierdo_ea_2021}. First, the centroid velocity residuals were folded and subtracted along the disk minor axis to remove axisymmetric features. Second, a 2D scan was performed to search for peak velocity residuals and obtain their locations in the folded map. Using these detected points, a K-means clustering algorithm then searched for coherent velocity perturbations within predefined radial and azimuthal bins \citep{MacQueen1967, Pedregosa2011}. We considered seven radial and ten azimuthal bins, which corresponds to a width of roughly one beam size, to identify clusters. The algorithm then subdivided the input residual points such that the center of each cluster is the closest to all points in the cluster, by iteratively minimizing the sum of squared distances from the data points to the center of the cluster. This led to irregularly spaced bin boundaries since the cluster centers are near the densest accumulations of points.

In Figure\,\ref{fig:polar_plot_clusters} we show the folded velocity residual map together with the detected peak velocity residuals (gray points). The location of the detected peak velocity residuals in azimuth and radius, within identified clusters, can be found in Fig.~\ref{fig:app_cluster_r_phi}. Clusters with high significance (those with peak velocity residuals larger than three times the variance in other clusters) are located within one radial and azimuthal bin shown in Fig.\,\ref{fig:polar_plot_clusters}. Taking the centers of the selected clusters allowed us to identify a localized perturbation at 0.28\arcsec{}\,±\,0.07\arcsec{} (${R=41\pm 10}$\,au) and ${\mathrm{PA}=280^\circ \pm\,2^\circ}$. The reported errors are the standard deviation of the peak residual point (R, $\phi$) locations within the selected clusters. The detection yields a cluster significance of $5.4\,\sigma_\phi$ in azimuth and $5.3\,\sigma_R$ in radius, where $\sigma$ represents the standard deviation of background cluster variances with a mean of $\sigma_\phi=0.034\,$km$^2$s$^{-2}$ and $\sigma_R=0.018\,$km$^2$s$^{-2}$ (see black crosses in Fig.\,\ref{fig:app_cluster_r_phi}). We note that a localized signature is robustly detected regardless of the amount of clusters defined, which we tested using 7-12 azimuthal or 5-9 radial clusters. We reported the clusters associated with the highest significance. Additionally, we note that there are other detections with lower significance at 0.65\arcsec{} (94\,au). This means that the radial extent of the prominent perturbation is roughly 0.40\arcsec{} (58\,au) and the global peak of the folded velocity residuals is at 0.39\arcsec{} (56\,au; as can be seen in the middle panel of Fig.\,\ref{fig:app_cluster_r_phi}). This analysis confirms the presence of a significant localized non-Keplerian feature, as identified visually in Fig.\,\ref{fig:centroid}\,(c), within the continuum ring.

\subsubsection{Spiral feature}
\label{sec:spiral}
\begin{figure}[t]
\centering
\includegraphics[width=\linewidth]{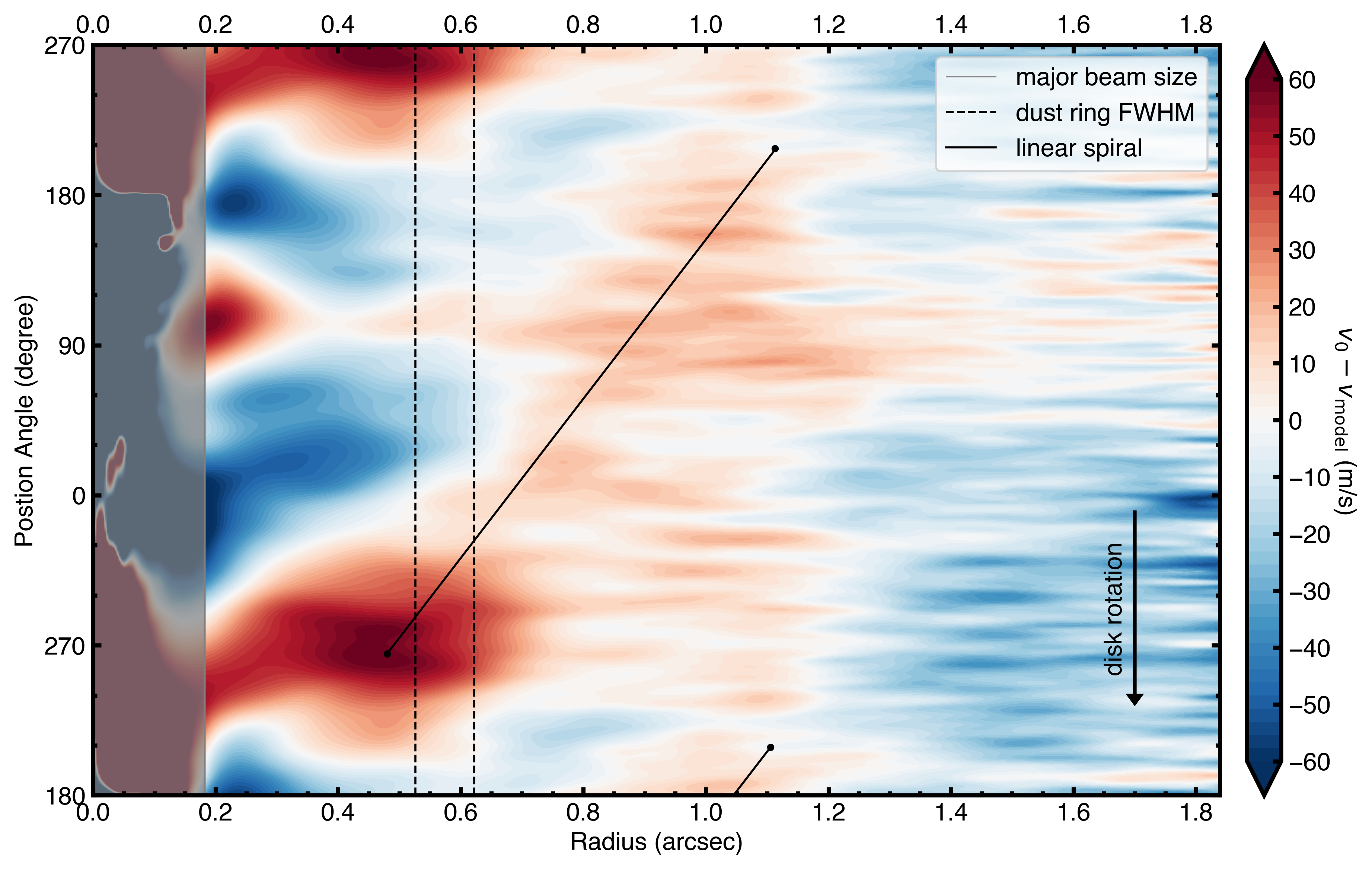}
\caption{Polar projection of the velocity residual map. The solid black line shows a linear spiral trace. The gray region indicates the masked area and the black dashed lines, the FWHM of the dust ring. The y-axis extends farther than 360$^\circ$ to enhance the visibility of the spiral.}
\label{fig:polar_map}
\vspace{-2.5mm}
\end{figure}

Figure\,\ref{fig:centroid}\,(c) also shows an extended arc-like positive residual feature,  beyond the dust continuum emission and covering nearly 300$^\circ$ in azimuth, that is more evident in the polar deprojection of the velocity residual map (Fig.\,\ref{fig:polar_map}).
To assess if this feature is a coherent structure, we used the \texttt{FilFinder} package \citep{Koch2015} implemented in \texttt{discminer} between 0.30\arcsec{} and 1.25\arcsec{} (43-180\,au; see Fig.\,\ref{fig:app_polar_filam_resid}). As indicated by the coherent filaments, the strong localized positive velocity residual discussed in Sect.\,\ref{sec:cent_resid} seems to be the starting point of a spiral tracing outward up to roughly 1.1\arcsec{} (159\,au). 
In Fig. \ref{fig:polar_map} we overlay an Archimedean (linear) spiral, prescribed by ${r_\mathrm{spiral}=a + b\,\phi_\mathrm{spiral}}$, using $\{a,b\}= \{0.48, 0.12\}$. Computing the pitch angles ${tan(\beta) = -(dr/d\phi)/r}$, we obtain values ranging from 14$^{\circ}$ to 6$^{\circ}$ over the spiral extent.

\subsubsection{A possible warp in the $^{12}$CO cavity}
\label{sec:warp}
An additional feature clearly evident from the velocity maps is the highly perturbed inner disk regions. As seen in the inset of the $v_\mathrm{0}$ line centroid map in Fig. \ref{fig:centroid}\,(a),  the iso-velocity lines show strong bending in the inner region ($\sim$3 beam sizes in diameter), indicative of non-Keplerian velocities. It likely traces a warp or a misaligned inner disk, as reported in \cite{Mayama2018}, \cite{Pinilla2018}, \cite{Sicilia2020}, and \cite{Ansdell2020} to explain the scattered-light shadows and variable photometry of J1604. Higher angular resolution deep gas observations are, however, needed to assess its morphology and kinematics.

\subsection{Deprojected velocity components}
\label{sec:deproj_velocities}
To understand the contributions from $v_{\phi}, v_{r}, v_{z}$, we produced three centroid residual maps for each velocity component, after deprojection (Eq.~\ref{eq:v_deproj}), assuming that ``all'' the velocities are azimuthal, radial, or vertical \citep[see Fig.\,\ref{fig:app_2D_depr_velocities};][]{Teague_ea_2022b}. The localized residual feature at the edge of the dust ring appears to trace variations in the vertical motions, $v_\mathrm{z}$, or in the rotational motions, $v_{\phi}$, or a combination of both. Radial perturbations can be ruled out since the feature is located close to the disk redshifted major axis (PA$_\mathrm{disk}$=258$^\circ$), where $v_{r,\mathrm{proj}}\approx0$. Assuming purely rotational velocities, this corresponds to perturbations as high as $\delta v_\phi\approx600$\,\msec ($\sim 0.4\cdot v_\mathrm{kepler}$) due to the low disk inclination.

As seen in Fig.~\ref{fig:centroid}\,(c), the spiral-like velocity residual feature does not change sign around the disk major or minor axes, which would occur for rotational, $v_{\phi}$, or radial, $v_{r}$, velocity perturbations, respectively (see Eq.~\ref{eq:v_deproj}). We are likely seeing vertical perturbations, which we are most sensitive to in a nearly face-on disk.

Figure \ref{fig:app_depr_velocities} shows the deprojected and azimuthally averaged radial profiles of each velocity component determined with \texttt{eddy}. For $v_{\phi}$, we observe super-Keplerian rotation from R$\sim$0.35\arcsec{}-0.70\arcsec{} (51-101\,au), peaking at 0.45\arcsec{} (65\,au), right beyond the dust continuum. The rotational velocities then sharply drop to sub-Keplerian in the inner disk regions. However, we stress that the azimuthally averaged velocities at the radial location of the strong localized perturbation (R$\sim0.3-0.6$\arcsec{},43-87\,au) are likely affected by the feature. We tentatively observe radial inflow inward of the CO intensity peak but with very large uncertainties on $v_r$. Finally, we mostly detect downward vertical motion of the disk within R$\sim$1.25\arcsec{} (181\,au).

\section{Discussion}
\label{sec:discussion}
\subsection{Origin of $v_0$ residuals}
\label{sec:discuss_planet}
In this Letter we report the detection of two main non-Keplerian features, in addition to highly perturbed gas velocities in the gas cavity: (1) a localized positive residual near the edge of the dust ring and (2) an extended spiral-like feature, possibly starting from (1). A variety of velocity residual features were detected in other systems, with a diverse range of inclinations \citep[e.g.,][]{Wolfer_ea_2022}. In the case of TW\,Hya, a similarly face-on disk, the detected perturbations are $\sim$40\,\msec \citep{Teague_ea_2022b}, which is lower than what is derived for (1), and in our case can account for 40\% of the local Keplerian velocity assuming that the perturbation is purely due to rotational velocities. This is also larger in the magnitude of deviation than the Doppler flip reported in the HD\,100546 transition disk \citep{Casassus_ea_2022}. These velocity residuals are often interpreted as tracing planet--disk interactions from massive companions \citep{PPVII_Pinte} that potentially carve out gaps. It is thus worth noting that our inferred planet location (${R=41\pm 10}$\,au) is close to the gap location in $^{13}$CO (J=2-1) at 37\,au reported in \cite{vanderMarel_ea_2021a}. Comparison with simulations \citep{Rabago2021,Izquierdo_ea_2022} or semi-analytical prescriptions \citep{Bollati2021} allowed us to estimate a possible planet mass from the velocity deviations. To this end, we considered Eq.\,14 from \cite{Yun_ea_2019}, which relates the change in rotational velocity, $\delta v_\phi$, to the planet mass, $M_p$, through 2D hydrodynamic simulations.

Since $\delta_V$ is the difference between the super- and sub-Keplerian peak, we considered the peak of the super-Keplerian motion, $(v_\phi/v_\mathrm{kep})/v_\mathrm{kep}$ (see Fig.~\ref{fig:app_depr_velocities}), as a lower limit for the dimensionless amplitude of the perturbed rotational velocity ($\delta_V^\mathrm{min}$=0.06), and its double ($\delta_V^\mathrm{max}=0.12$) as an upper limit. Assuming $(H/R)_p=0.1$ at the planet location ($R=41\,$au), we estimate the planet mass to roughly be in the range $M_p\approx(1.6-2.9)M_\mathrm{jup}\,(\alpha/10^{-3})^{0.5}$.

The extended spiral-like feature appears to be related to the significant localized velocity residual. Due to its low pitch angles and the consistent positive velocity residuals, we speculate that the spiral is caused by buoyancy resonances excited by a massive planet located within the dust ring. Indeed, in contrast with Lindblad  spirals, buoyancy spirals are shown to exhibit a tightly wound morphology with predominantly vertical motions \citep{Bae_ea_2021}. Such a spiral has also been suggested in TW\,Hya \citep{Teague_ea_2019a}, whose radial extent is similar to the one reported here.
Interestingly, \cite{Wolfer_ea_2022} report a tentative arc feature in J1604,  at $R\sim1.0$\arcsec{} and ranging from PA$\approx160-200^{\circ}$, probed by the kinematics of the $^{12}$CO (J=3-2) line emission, that partly coincides with the spiral-like feature that we detect. Additional observations in optically thin tracers would be very useful to assess its nature.

\subsection{Kinematic perturbations due to shadows}
\label{sec:discuss_shadows}
The localized residual feature seems to roughly align with the shadows detected in scattered light. Comparing Figs.\,\ref{fig:intensities}\,(c) and \ref{fig:centroid}\,(c), positive and negative $v_0$ residuals broadly align with cold and hot regions in the brightness temperature of $^{12}$CO, respectively (see also Fig.\,\ref{fig:app_contour_resid}). In particular, the  orientation of the significant localized velocity perturbation coincides with the western shadow. Such a shadow can cool down the disk material and possibly induce a local drop in pressure support and therefore impact the gas velocity.
In this section we estimate whether the detected velocity perturbations could be caused by azimuthal variations in temperature.  We related the azimuthal change in temperature, $\Delta T_\phi$, to variations in rotational velocity, $\Delta v_\phi$, by solving for the Navier-Stokes equation in cylindrical coordinates. Following the derivation in Appendix \ref{app:nav_st_deriv}, we obtain
\begin{equation}
    \frac{\Delta v_\phi}{v_\mathrm{kep}} \approx \left(\frac{H}{R}\right)^2 \frac{\Delta T_\phi}{T},
\end{equation}
with $H/R$ the disk aspect ratio, $T$ the $^{12}$CO brightness temperature tracing the gas temperature (as $^{12}$CO is optically thick), and $v_\mathrm{kep}$ the Keplerian velocity. Evaluating this equation for a {large} $H/R=0.2$ along a radially averaged annulus centered at $R=({0.39\pm0.07}$)\arcsec{}, where we experience the strongest azimuthal changes in the $^{12}$CO brightness temperature of up to $\Delta T\approx15$K over $T=60$K, we estimate the change in azimuthal velocity to be a mere $\delta v_\phi/v_\mathrm{kep} \approx 1\%$. Hence, the shadows cannot be solely responsible for the localized velocity residual feature. In addition, as the shadows are nearly symmetric, we would expect a similar feature opposite to its east.

\cite{Montesinos_ea_2016} investigated the development of spirals due to pressure gradients caused by temperature differences between obscured and illuminated regions. In their simulations, symmetric shadows always form two-armed spirals. However, they only develop for massive ($0.25\,M_\star$) and/or strongly illuminated disks (100\,$L_\odot$), which does not seem to be the case for J1604 \citep[$\sim0.02\,M_\odot$ \& 0.7\,$L_\odot$,][]{Manara_ea_2020}, where we also only observe one spiral. Additionally, we would also expect such spiral features to appear in the brightness temperature residuals (Fig.~\ref{fig:intensities}\,c). It is therefore unlikely that shadows are responsible for the extended spiral-like velocity residual feature.

\subsection{A warped or misaligned inner disk?}
\label{sec:discuss_warp}
The bending of the iso-velocity curves that we observe in the inset of Fig.~\ref{fig:centroid}\,(a) is reminiscent of a warped or (broken) misaligned inner disk \citep{Juhasz2017,Facchini_ea_2018}. However, as the inner disk is unresolved in our observations, the warp morphology cannot be derived. We note that radial inflows are also degenerate in appearance with warps, as shown by \cite{Rosenfeld_ea_2014}, and that our observations do not allow us to be conclusive regarding the origin of the disturbed kinematics in the innermost disk. We attempted to infer if the position angle or inclination vary with radius by fitting the innermost disk only (R$\leq$0.5\arcsec{}) with \texttt{eddy} and considering a fixed stellar mass, but we did not find any significant variations relative to our best-fit values. We therefore constrain the warp to be confined within one beam size ($\sim$0.18\arcsec{}, i.e., 26\,au) from the center.
We note that we obtain a 5\,\% higher dynamical mass of the system when fitting for $M_\star$ while masking the innermost beam size in radius, an effect predicted by hydrodynamical simulations of warps \citep{Young_ea_2022}.

While our observations cannot provide a full picture of the system due to a limited angular resolution, the very low mass accretion rate and near-infrared excess \citep{Sicilia2020}, as well as the gas cavity in $^{12}$CO with non-Keplerian velocities, suggest the presence of an additional, very massive (possibly stellar) companion within the inner $\sim$0.25\arcsec{} ($\sim$35\,au). Such a companion would need to be on an inclined (nearly polar) orbit to misalign the inner disk \citep{Zhu2019}, which would then lead to the shadows \citep{Nealon2019} and variable extinction events seen in the light curves \citep{Ansdell2020, Sicilia2020}. It would not, however, explain the strong localized velocity residual feature that we report, which we speculate traces a planetary-mass object located at the edge of the dust continuum. Detailed modeling of the system is thus needed to assess the need for an additional companion. An interesting comparison is the CS\,Cha spectro-binary system (separation of $\sim$7\,au), which shows a similar dust continuum and gas emission at a similarly low inclination but no departure from Keplerian rotation in the $^{12}$CO kinematics \citep{Kurtovic_ea_2022}.

\section{Conclusions}
\label{sec:conclusion}
In this Letter we present new ALMA observations of the 1.3\,mm dust continuum and the $^{12}$CO (J=2-1) line emission from the transition disk around RXJ1604.3–2130 A. The dust continuum shows a large cavity enclosing a smaller $^{12}$CO cavity. Azimuthal brightness variations in the $^{12}$CO line and dust continuum are broadly aligned with shadows detected in scattered light \citep{Pinilla2018}. Using the \texttt{discminer} package \citep{Izquierdo_ea_2021}, we modeled the channel-by-channel line emission and calculated the line-of-sight velocity maps.
We report the detection of a coherent, localized non-Keplerian feature at ${R=(41\pm10)\,\mathrm{au}}$ and ${PA=280^\circ\pm 2^\circ}$, that is, within the continuum ring. 
While broadly aligned with the scattered-light shadows, the localized non-Keplerian feature cannot be due to changes in temperature. Instead, we interpret the kinematical perturbation as tracing the presence of a massive companion of $M_p\approx(1.6-2.9)\,M_\mathrm{jup}$. We also detect a tightly wound spiral that extends over $300^\circ$ in azimuth, possibly connected to the localized feature and caused by buoyancy resonances driven by planet--disk interactions. Bending of the iso-velocity contours within the gas cavity indicates a highly perturbed inner region, possibly related to the presence of a misaligned inner disk. However, as the putative planet at $\sim$41 au cannot explain the gas cavity, the low accretion rate, or the misaligned inner disk, we speculate that another massive companion, likely on an inclined orbit, shapes the inner $\sim$0.25\arcsec{}($\sim$35\,au).

\begin{acknowledgements}
We would like to thank the anonymous referee for the constructive feedback, as well as Clement Baruteau, Kees Dullemond, Guillaume Laibe and Andrew Winter for helpful discussions. This Letter makes use of the following ALMA data: ADS/JAO.ALMA\#2017.A.01255.S and ADS/JAO.ALMA\#2015.1.00964.S. ALMA is a partnership of ESO (representing its member states), NSF (USA), and NINS (Japan), together with NRC (Canada),  NSC and ASIAA (Taiwan), and KASI (Republic of Korea), in cooperation with the Republic of Chile. The Joint ALMA Observatory is operated by ESO, AUI/NRAO, and NAOJ. This project has received funding from the European Research Council (ERC) under the European Union’s Horizon 2020 research and innovation programme (PROTOPLANETS, grant agreement No. 101002188). L.P. gratefully acknowledges support by the ANID BASAL projects ACE210002 and FB210003, and by ANID, -- Millennium Science Initiative Program -- NCN19\_171. Software: CARTA \citep{Comrie_2021_carta}, CASA \citep{McMullin2007}, Discminer \citep{Izquierdo_ea_2021}, Eddy \citep{Teague_2019_eddy}, FilFinder \citep{Koch2015}, GoFish \citep{Teague_2019_gofish}, Matplotlib \citep{Hunter_mpl}, Numpy \citep{vanderWalt_np}, Scipy \citep{Virtanen_scipy}.
\end{acknowledgements}

\bibpunct{(}{)}{;}{a}{}{,} 
\bibliographystyle{aa} 
\bibliography{mybib.bib} 

\begin{thebibliography}{66}
\expandafter\ifx\csname natexlab\endcsname\relax\def\natexlab#1{#1}\fi

\bibitem[{{Andrews} {et~al.}(2018){Andrews}, {Huang}, {P{\'e}rez}, {Isella},
  {Dullemond}, {Kurtovic}, {Guzm{\'a}n}, {Carpenter}, {Wilner}, {Zhang}, {Zhu},
  {Birnstiel}, {Bai}, {Benisty}, {Hughes}, {{\"O}berg}, \&
  {Ricci}}]{Andrews_ea_2018}
{Andrews}, S.~M., {Huang}, J., {P{\'e}rez}, L.~M., {et~al.} 2018, \apjl, 869,
  L41

\bibitem[{{Ansdell} {et~al.}(2020){Ansdell}, {Gaidos}, {Hedges}, {Tazzari},
  {Kraus}, {Wyatt}, {Kennedy}, {Williams}, {Mann}, {Angelo}, {D{\^u}chene},
  {Mamajek}, {Carpenter}, {Esplin}, \& {Rizzuto}}]{Ansdell2020}
{Ansdell}, M., {Gaidos}, E., {Hedges}, C., {et~al.} 2020, \mnras, 492, 572

\bibitem[{{Bae} {et~al.}(2022{\natexlab{a}}){Bae}, {Isella}, {Zhu}, {Martin},
  {Okuzumi}, \& {Suriano}}]{Bae_PPVII}
{Bae}, J., {Isella}, A., {Zhu}, Z., {et~al.} 2022{\natexlab{a}}, arXiv
  e-prints, arXiv:2210.13314

\bibitem[{{Bae} {et~al.}(2022{\natexlab{b}}){Bae}, {Teague}, {Andrews},
  {Benisty}, {Facchini}, {Galloway-Sprietsma}, {Loomis}, {Aikawa},
  {Alarc{\'o}n}, {Bergin}, {Bergner}, {Booth}, {Cataldi}, {Cleeves}, {Czekala},
  {Guzm{\'a}n}, {Huang}, {Ilee}, {Kurtovic}, {Law}, {Gal}, {Liu}, {Long},
  {M{\'e}nard}, {{\"O}berg}, {P{\'e}rez}, {Qi}, {Schwarz}, {Sierra}, {Walsh},
  {Wilner}, \& {Zhang}}]{Bae_ea_2022}
{Bae}, J., {Teague}, R., {Andrews}, S.~M., {et~al.} 2022{\natexlab{b}}, \apjl,
  934, L20

\bibitem[{{Bae} {et~al.}(2021){Bae}, {Teague}, \& {Zhu}}]{Bae_ea_2021}
{Bae}, J., {Teague}, R., \& {Zhu}, Z. 2021, \apj, 912, 56

\bibitem[{{Barenfeld} {et~al.}(2016){Barenfeld}, {Carpenter}, {Ricci}, \&
  {Isella}}]{Barenfeld2016}
{Barenfeld}, S.~A., {Carpenter}, J.~M., {Ricci}, L., \& {Isella}, A. 2016,
  \apj, 827, 142

\bibitem[{{Benisty} {et~al.}(2022){Benisty}, {Dominik}, {Follette}, {Garufi},
  {Ginski}, {Hashimoto}, {Keppler}, {Kley}, \& {Monnier}}]{PPVII_Benisty}
{Benisty}, M., {Dominik}, C., {Follette}, K., {et~al.} 2022, arXiv e-prints,
  arXiv:2203.09991

\bibitem[{{Bollati} {et~al.}(2021){Bollati}, {Lodato}, {Price}, \&
  {Pinte}}]{Bollati2021}
{Bollati}, F., {Lodato}, G., {Price}, D.~J., \& {Pinte}, C. 2021, \mnras, 504,
  5444

\bibitem[{{Calcino} {et~al.}(2022){Calcino}, {Hilder}, {Price}, {Pinte},
  {Bollati}, {Lodato}, \& {Norfolk}}]{Calcino2022}
{Calcino}, J., {Hilder}, T., {Price}, D.~J., {et~al.} 2022, \apjl, 929, L25

\bibitem[{{Casassus} {et~al.}(2022){Casassus}, {C{\'a}rcamo}, {Hales}, {Weber},
  \& {Dent}}]{Casassus_ea_2022}
{Casassus}, S., {C{\'a}rcamo}, M., {Hales}, A., {Weber}, P., \& {Dent}, B.
  2022, \apjl, 933, L4

\bibitem[{{Casassus} \& {P{\'e}rez}(2019)}]{Casassus_Perez_2019}
{Casassus}, S. \& {P{\'e}rez}, S. 2019, \apjl, 883, L41

\bibitem[{{Comrie} {et~al.}(2021){Comrie}, {Wang}, {Hsu}, {Moraghan}, {Harris},
  {Pang}, {Pi{\'n}ska}, {Chiang}, {Chang}, {Hwang}, {Jan}, {Lin}, \&
  {Simmonds}}]{Comrie_2021_carta}
{Comrie}, A., {Wang}, K.-S., {Hsu}, S.-C., {et~al.} 2021, {CARTA: The Cube
  Analysis and Rendering Tool for Astronomy}, Zenodo

\bibitem[{{Czekala} {et~al.}(2021){Czekala}, {Loomis}, {Teague}, {Booth},
  {Huang}, {Cataldi}, {Ilee}, {Law}, {Walsh}, {Bosman}, {Guzm{\'a}n}, {Gal},
  {{\"O}berg}, {Yamato}, {Aikawa}, {Andrews}, {Bae}, {Bergin}, {Bergner},
  {Cleeves}, {Kurtovic}, {M{\'e}nard}, {Nomura}, {P{\'e}rez}, {Qi}, {Schwarz},
  {Tsukagoshi}, {Waggoner}, {Wilner}, \& {Zhang}}]{czekala2021}
{Czekala}, I., {Loomis}, R.~A., {Teague}, R., {et~al.} 2021, \apjs, 257, 2

\bibitem[{{Dong} {et~al.}(2017){Dong}, {van der Marel}, {Hashimoto}, {Chiang},
  {Akiyama}, {Liu}, {Muto}, {Knapp}, {Tsukagoshi}, {Brown}, {Bruderer},
  {Koyamatsu}, {Kudo}, {Ohashi}, {Rich}, {Satoshi}, {Takami}, {Wisniewski},
  {Yang}, {Zhu}, \& {Tamura}}]{Dong2017}
{Dong}, R., {van der Marel}, N., {Hashimoto}, J., {et~al.} 2017, \apj, 836, 201

\bibitem[{{Facchini} {et~al.}(2018{\natexlab{a}}){Facchini}, {Juh{\'a}sz}, \&
  {Lodato}}]{Facchini_ea_2018}
{Facchini}, S., {Juh{\'a}sz}, A., \& {Lodato}, G. 2018{\natexlab{a}}, \mnras,
  473, 4459

\bibitem[{{Facchini} {et~al.}(2018{\natexlab{b}}){Facchini}, {Pinilla}, {van
  Dishoeck}, \& {de Juan Ovelar}}]{Facchini2018}
{Facchini}, S., {Pinilla}, P., {van Dishoeck}, E.~F., \& {de Juan Ovelar}, M.
  2018{\natexlab{b}}, \aap, 612, A104

\bibitem[{{Foreman-Mackey} {et~al.}(2013){Foreman-Mackey}, {Hogg}, {Lang}, \&
  {Goodman}}]{emcee2013}
{Foreman-Mackey}, D., {Hogg}, D.~W., {Lang}, D., \& {Goodman}, J. 2013, \pasp,
  125, 306

\bibitem[{{Gaia Collaboration} {et~al.}(2022){Gaia Collaboration}, {Vallenari},
  {Brown}, {Prusti}, {de Bruijne}, {Arenou}, {Babusiaux}, {Biermann},
  {Creevey}, {Ducourant}, {Evans}, {Eyer}, {Guerra}, {Hutton}, {Jordi},
  {Klioner}, {Lammers}, {Lindegren}, {Luri}, {Mignard}, {Panem}, {Pourbaix},
  {Randich}, {Sartoretti}, {Soubiran}, {Tanga}, {Walton}, {Bailer-Jones},
  {Bastian}, {Drimmel}, {Jansen}, {Katz}, {Lattanzi}, {van Leeuwen}, {Bakker},
  {Cacciari}, {Casta{\~n}eda}, {De Angeli}, {Fabricius}, {Fouesneau},
  {Fr{\'e}mat}, {Galluccio}, {Guerrier}, {Heiter}, {Masana}, {Messineo},
  {Mowlavi}, {Nicolas}, {Nienartowicz}, {Pailler}, {Panuzzo}, {Riclet}, {Roux},
  {Seabroke}, {Sordo{\o}rcit}, {Th{\'e}venin}, {Gracia-Abril}, {Portell},
  {Teyssier}, {Altmann}, {Andrae}, {Audard}, {Bellas-Velidis}, {Benson},
  {Berthier}, {Blomme}, {Burgess}, {Busonero}, {Busso}, {C{\'a}novas}, {Carry},
  {Cellino}, {Cheek}, {Clementini}, {Damerdji}, {Davidson}, {de Teodoro},
  {Nu{\~n}ez Campos}, {Delchambre}, {Dell'Oro}, {Esquej},
  {Fern{\'a}ndez-Hern{\'a}ndez}, {Fraile}, {Garabato}, {Garc{\'\i}a-Lario},
  {Gosset}, {Haigron}, {Halbwachs}, {Hambly}, {Harrison}, {Hern{\'a}ndez},
  {Hestroffer}, {Hodgkin}, {Holl}, {Jan{\ss}en}, {Jevardat de Fombelle},
  {Jordan}, {Krone-Martins}, {Lanzafame}, {L{\"o}ffler}, {Marchal}, {Marrese},
  {Moitinho}, {Muinonen}, {Osborne}, {Pancino}, {Pauwels}, {Recio-Blanco},
  {Reyl{\'e}}, {Riello}, {Rimoldini}, {Roegiers}, {Rybizki}, {Sarro}, {Siopis},
  {Smith}, {Sozzetti}, {Utrilla}, {van Leeuwen}, {Abbas}, {{\'A}brah{\'a}m},
  {Abreu Aramburu}, {Aerts}, {Aguado}, {Ajaj}, {Aldea-Montero}, {Altavilla},
  {{\'A}lvarez}, {Alves}, {Anders}, {Anderson}, {Anglada Varela}, {Antoja},
  {Baines}, {Baker}, {Balaguer-N{\'u}{\~n}ez}, {Balbinot}, {Balog}, {Barache},
  {Barbato}, {Barros}, {Barstow}, {Bartolom{\'e}}, {Bassilana}, {Bauchet},
  {Becciani}, {Bellazzini}, {Berihuete}, {Bernet}, {Bertone}, {Bianchi},
  {Binnenfeld}, {Blanco-Cuaresma}, {Blazere}, {Boch}, {Bombrun}, {Bossini},
  {Bouquillon}, {Bragaglia}, {Bramante}, {Breedt}, {Bressan}, {Brouillet},
  {Brugaletta}, {Bucciarelli}, {Burlacu}, {Butkevich}, {Buzzi}, {Caffau},
  {Cancelliere}, {Cantat-Gaudin}, {Carballo}, {Carlucci}, {Carnerero},
  {Carrasco}, {Casamiquela}, {Castellani}, {Castro-Ginard}, {Chaoul},
  {Charlot}, {Chemin}, {Chiaramida}, {Chiavassa}, {Chornay}, {Comoretto},
  {Contursi}, {Cooper}, {Cornez}, {Cowell}, {Crifo}, {Cropper}, {Crosta},
  {Crowley}, {Dafonte}, {Dapergolas}, {David}, {David}, {de Laverny}, {De
  Luise}, {De March}, {De Ridder}, {de Souza}, {de Torres}, {del Peloso}, {del
  Pozo}, {Delbo}, {Delgado}, {Delisle}, {Demouchy}, {Dharmawardena}, {Di
  Matteo}, {Diakite}, {Diener}, {Distefano}, {Dolding}, {Edvardsson}, {Enke},
  {Fabre}, {Fabrizio}, {Faigler}, {Fedorets}, {Fernique}, {Fienga}, {Figueras},
  {Fournier}, {Fouron}, {Fragkoudi}, {Gai}, {Garcia-Gutierrez},
  {Garcia-Reinaldos}, {Garc{\'\i}a-Torres}, {Garofalo}, {Gavel}, {Gavras},
  {Gerlach}, {Geyer}, {Giacobbe}, {Gilmore}, {Girona}, {Giuffrida}, {Gomel},
  {Gomez}, {Gonz{\'a}lez-N{\'u}{\~n}ez}, {Gonz{\'a}lez-Santamar{\'\i}a},
  {Gonz{\'a}lez-Vidal}, {Granvik}, {Guillout}, {Guiraud},
  {Guti{\'e}rrez-S{\'a}nchez}, {Guy}, {Hatzidimitriou}, {Hauser}, {Haywood},
  {Helmer}, {Helmi}, {Sarmiento}, {Hidalgo}, {Hilger}, {H{\l}adczuk}, {Hobbs},
  {Holland}, {Huckle}, {Jardine}, {Jasniewicz}, {Jean-Antoine Piccolo},
  {Jim{\'e}nez-Arranz}, {Jorissen}, {Juaristi Campillo}, {Julbe}, {Karbevska},
  {Kervella}, {Khanna}, {Kontizas}, {Kordopatis}, {Korn}, {K{\'o}sp{\'a}l},
  {Kostrzewa-Rutkowska}, {Kruszy{\'n}ska}, {Kun}, {Laizeau}, {Lambert},
  {Lanza}, {Lasne}, {Le Campion}, {Lebreton}, {Lebzelter}, {Leccia}, {Leclerc},
  {Lecoeur-Taibi}, {Liao}, {Licata}, {Lindstr{\o}m}, {Lister}, {Livanou},
  {Lobel}, {Lorca}, {Loup}, {Madrero Pardo}, {Magdaleno Romeo}, {Managau},
  {Mann}, {Manteiga}, {Marchant}, {Marconi}, {Marcos}, {Marcos Santos},
  {Mar{\'\i}n Pina}, {Marinoni}, {Marocco}, {Marshall}, {Polo},
  {Mart{\'\i}n-Fleitas}, {Marton}, {Mary}, {Masip}, {Massari},
  {Mastrobuono-Battisti}, {Mazeh}, {McMillan}, {Messina}, {Michalik}, {Millar},
  {Mints}, {Molina}, {Molinaro}, {Moln{\'a}r}, {Monari}, {Mongui{\'o}},
  {Montegriffo}, {Montero}, {Mor}, {Mora}, {Morbidelli}, {Morel}, {Morris},
  {Muraveva}, {Murphy}, {Musella}, {Nagy}, {Noval}, {Oca{\~n}a}, {Ogden},
  {Ordenovic}, {Osinde}, {Pagani}, {Pagano}, {Palaversa}, {Palicio},
  {Pallas-Quintela}, {Panahi}, {Payne-Wardenaar}, {Pe{\~n}alosa Esteller},
  {Penttil{\"a}}, {Pichon}, {Piersimoni}, {Pineau}, {Plachy}, {Plum}, {Poggio},
  {Pr{\v{s}}a}, {Pulone}, {Racero}, {Ragaini}, {Rainer}, {Raiteri}, {Rambaux},
  {Ramos}, {Ramos-Lerate}, {Re Fiorentin}, {Regibo}, {Richards}, {Rios Diaz},
  {Ripepi}, {Riva}, {Rix}, {Rixon}, {Robichon}, {Robin}, {Robin}, {Roelens},
  {Rogues}, {Rohrbasser}, {Romero-G{\'o}mez}, {Rowell}, {Royer}, {Ruz Mieres},
  {Rybicki}, {Sadowski}, {S{\'a}ez N{\'u}{\~n}ez}, {Sagrist{\`a} Sell{\'e}s},
  {Sahlmann}, {Salguero}, {Samaras}, {Sanchez Gimenez}, {Sanna},
  {Santove{\~n}a}, {Sarasso}, {Schultheis}, {Sciacca}, {Segol}, {Segovia},
  {S{\'e}gransan}, {Semeux}, {Shahaf}, {Siddiqui}, {Siebert}, {Siltala},
  {Silvelo}, {Slezak}, {Slezak}, {Smart}, {Snaith}, {Solano}, {Solitro},
  {Souami}, {Souchay}, {Spagna}, {Spina}, {Spoto}, {Steele},
  {Steidelm{\"u}ller}, {Stephenson}, {S{\"u}veges}, {Surdej}, {Szabados},
  {Szegedi-Elek}, {Taris}, {Taylo}, {Teixeira}, {Tolomei}, {Tonello}, {Torra},
  {Torra}, {Torralba Elipe}, {Trabucchi}, {Tsounis}, {Turon}, {Ulla}, {Unger},
  {Vaillant}, {van Dillen}, {van Reeven}, {Vanel}, {Vecchiato}, {Viala},
  {Vicente}, {Voutsinas}, {Weiler}, {Wevers}, {Wyrzykowski}, {Yoldas}, {Yvard},
  {Zhao}, {Zorec}, {Zucker}, \& {Zwitter}}]{Gaia_dr3_2022}
{Gaia Collaboration}, {Vallenari}, A., {Brown}, A.~G.~A., {et~al.} 2022, arXiv
  e-prints, arXiv:2208.00211

\bibitem[{Hunter(2007)}]{Hunter_mpl}
Hunter, J.~D. 2007, Computing in Science Engineering, 9, 90

\bibitem[{{Izquierdo} {et~al.}(2022){Izquierdo}, {Facchini}, {Rosotti}, {van
  Dishoeck}, \& {Testi}}]{Izquierdo_ea_2022}
{Izquierdo}, A.~F., {Facchini}, S., {Rosotti}, G.~P., {van Dishoeck}, E.~F., \&
  {Testi}, L. 2022, \apj, 928, 2

\bibitem[{{Izquierdo} {et~al.}(2021){Izquierdo}, {Testi}, {Facchini},
  {Rosotti}, \& {van Dishoeck}}]{Izquierdo_ea_2021}
{Izquierdo}, A.~F., {Testi}, L., {Facchini}, S., {Rosotti}, G.~P., \& {van
  Dishoeck}, E.~F. 2021, \aap, 650, A179

\bibitem[{{Jorsater} \& {van Moorsel}(1995)}]{jvm1995}
{Jorsater}, S. \& {van Moorsel}, G.~A. 1995, \aj, 110, 2037

\bibitem[{{Juh{\'a}sz} \& {Facchini}(2017)}]{Juhasz2017}
{Juh{\'a}sz}, A. \& {Facchini}, S. 2017, \mnras, 466, 4053

\bibitem[{{Keppler} {et~al.}(2019){Keppler}, {Teague}, {Bae}, {Benisty},
  {Henning}, {van Boekel}, {Chapillon}, {Pinilla}, {Williams}, {Bertrang},
  {Facchini}, {Flock}, {Ginski}, {Juhasz}, {Klahr}, {Liu}, {M{\"u}ller},
  {P{\'e}rez}, {Pohl}, {Rosotti}, {Samland}, \& {Semenov}}]{Keppler_ea_2019}
{Keppler}, M., {Teague}, R., {Bae}, J., {et~al.} 2019, \aap, 625, A118

\bibitem[{{Koch} \& {Rosolowsky}(2015)}]{Koch2015}
{Koch}, E.~W. \& {Rosolowsky}, E.~W. 2015, \mnras, 452, 3435

\bibitem[{{K{\"o}hler} {et~al.}(2000){K{\"o}hler}, {Kunkel}, {Leinert}, \&
  {Zinnecker}}]{Koehler_ea_2000}
{K{\"o}hler}, R., {Kunkel}, M., {Leinert}, C., \& {Zinnecker}, H. 2000, \aap,
  356, 541

\bibitem[{{Kostov} {et~al.}(2016){Kostov}, {Orosz}, {Welsh}, {Doyle},
  {Fabrycky}, {Haghighipour}, {Quarles}, {Short}, {Cochran}, {Endl}, {Ford},
  {Gregorio}, {Hinse}, {Isaacson}, {Jenkins}, {Jensen}, {Kane}, {Kull},
  {Latham}, {Lissauer}, {Marcy}, {Mazeh}, {M{\"u}ller}, {Pepper}, {Quinn},
  {Ragozzine}, {Shporer}, {Steffen}, {Torres}, {Windmiller}, \&
  {Borucki}}]{Kostov2016}
{Kostov}, V.~B., {Orosz}, J.~A., {Welsh}, W.~F., {et~al.} 2016, \apj, 827, 86

\bibitem[{{Kurtovic} {et~al.}(2022){Kurtovic}, {Pinilla}, {Penzlin}, {Benisty},
  {P{\'e}rez}, {Ginski}, {Isella}, {Kley}, {Menard}, {P{\'e}rez}, \&
  {Bayo}}]{Kurtovic_ea_2022}
{Kurtovic}, N.~T., {Pinilla}, P., {Penzlin}, A. B.~T., {et~al.} 2022, \aap,
  664, A151

\bibitem[{{Law} {et~al.}(2021){Law}, {Teague}, {Loomis}, {Bae}, {{\"O}berg},
  {Czekala}, {Andrews}, {Aikawa}, {Alarc{\'o}n}, {Bergin}, {Bergner}, {Booth},
  {Bosman}, {Calahan}, {Cataldi}, {Cleeves}, {Furuya}, {Guzm{\'a}n}, {Huang},
  {Ilee}, {Le Gal}, {Liu}, {Long}, {M{\'e}nard}, {Nomura}, {P{\'e}rez}, {Qi},
  {Schwarz}, {Soto}, {Tsukagoshi}, {Yamato}, {van't Hoff}, {Walsh}, {Wilner},
  \& {Zhang}}]{Law_ea_2021b}
{Law}, C.~J., {Teague}, R., {Loomis}, R.~A., {et~al.} 2021, \apjs, 257, 4

\bibitem[{{Long} {et~al.}(2018){Long}, {Pinilla}, {Herczeg}, {Harsono},
  {Dipierro}, {Pascucci}, {Hendler}, {Tazzari}, {Ragusa}, {Salyk}, {Edwards},
  {Lodato}, {van de Plas}, {Johnstone}, {Liu}, {Boehler}, {Cabrit}, {Manara},
  {Menard}, {Mulders}, {Nisini}, {Fischer}, {Rigliaco}, {Banzatti}, {Avenhaus},
  \& {Gully-Santiago}}]{Long_ea_2018}
{Long}, F., {Pinilla}, P., {Herczeg}, G.~J., {et~al.} 2018, \apj, 869, 17

\bibitem[{MacQueen(1967)}]{MacQueen1967}
MacQueen, J. 1967, in 5th Berkeley Symp. Math. Statist. Probability, 281--297

\bibitem[{{Manara} {et~al.}(2020){Manara}, {Natta}, {Rosotti}, {Alcal{\'a}},
  {Nisini}, {Lodato}, {Testi}, {Pascucci}, {Hillenbrand}, {Carpenter},
  {Scholz}, {Fedele}, {Frasca}, {Mulders}, {Rigliaco}, {Scardoni}, \&
  {Zari}}]{Manara_ea_2020}
{Manara}, C.~F., {Natta}, A., {Rosotti}, G.~P., {et~al.} 2020, \aap, 639, A58

\bibitem[{{Mayama} {et~al.}(2018){Mayama}, {Akiyama}, {Pani{\'c}}, {Miley},
  {Tsukagoshi}, {Muto}, {Dong}, {de Leon}, {Mizuki}, {Oh}, {Hashimoto}, {Sai},
  {Currie}, {Takami}, {Grady}, {Hayashi}, {Tamura}, \& {Inutsuka}}]{Mayama2018}
{Mayama}, S., {Akiyama}, E., {Pani{\'c}}, O., {et~al.} 2018, \apjl, 868, L3

\bibitem[{{McMullin} {et~al.}(2007){McMullin}, {Waters}, {Schiebel}, {Young},
  \& {Golap}}]{McMullin2007}
{McMullin}, J.~P., {Waters}, B., {Schiebel}, D., {Young}, W., \& {Golap}, K.
  2007, in Astronomical Society of the Pacific Conference Series, Vol. 376,
  Astronomical Data Analysis Software and Systems XVI, ed. R.~A. {Shaw},
  F.~{Hill}, \& D.~J. {Bell}, 127

\bibitem[{{Montesinos} {et~al.}(2016){Montesinos}, {Perez}, {Casassus},
  {Marino}, {Cuadra}, \& {Christiaens}}]{Montesinos_ea_2016}
{Montesinos}, M., {Perez}, S., {Casassus}, S., {et~al.} 2016, \apjl, 823, L8

\bibitem[{{Nealon} {et~al.}(2019){Nealon}, {Pinte}, {Alexander}, {Mentiplay},
  \& {Dipierro}}]{Nealon2019}
{Nealon}, R., {Pinte}, C., {Alexander}, R., {Mentiplay}, D., \& {Dipierro}, G.
  2019, \mnras, 484, 4951

\bibitem[{{Norfolk} {et~al.}(2022){Norfolk}, {Pinte}, {Calcino}, {Hammond},
  {van der Marel}, {Price}, {Maddison}, {Christiaens}, {Gonzalez}, {Blakely},
  {Rosotti}, \& {Ginski}}]{Norfolk2022}
{Norfolk}, B.~J., {Pinte}, C., {Calcino}, J., {et~al.} 2022, \apjl, 936, L4

\bibitem[{Pedregosa {et~al.}(2011)Pedregosa, Varoquaux, Gramfort, Michel,
  Thirion, Grisel, Blondel, Prettenhofer, Weiss, Dubourg,
  {et~al.}}]{Pedregosa2011}
Pedregosa, F., Varoquaux, G., Gramfort, A., {et~al.} 2011, the Journal of
  machine Learning research, 12, 2825

\bibitem[{{Perez} {et~al.}(2015){Perez}, {Dunhill}, {Casassus}, {Roman},
  {Szul{\'a}gyi}, {Flores}, {Marino}, \& {Montesinos}}]{Perez_ea_2015}
{Perez}, S., {Dunhill}, A., {Casassus}, S., {et~al.} 2015, \apjl, 811, L5

\bibitem[{{Pinilla} {et~al.}(2018){Pinilla}, {Benisty}, {de Boer}, {Manara},
  {Bouvier}, {Dominik}, {Ginski}, {Loomis}, \& {Sicilia Aguilar}}]{Pinilla2018}
{Pinilla}, P., {Benisty}, M., {de Boer}, J., {et~al.} 2018, \apj, 868, 85

\bibitem[{{Pinilla} {et~al.}(2015){Pinilla}, {de Boer}, {Benisty},
  {Juh{\'a}sz}, {de Juan Ovelar}, {Dominik}, {Avenhaus}, {Birnstiel}, {Girard},
  {Huelamo}, {Isella}, \& {Milli}}]{Pinilla2015}
{Pinilla}, P., {de Boer}, J., {Benisty}, M., {et~al.} 2015, \aap, 584, L4

\bibitem[{{Pinte} {et~al.}(2022){Pinte}, {Teague}, {Flaherty}, {Hall},
  {Facchini}, \& {Casassus}}]{PPVII_Pinte}
{Pinte}, C., {Teague}, R., {Flaherty}, K., {et~al.} 2022, arXiv e-prints,
  arXiv:2203.09528

\bibitem[{{Rabago} \& {Zhu}(2021)}]{Rabago2021}
{Rabago}, I. \& {Zhu}, Z. 2021, \mnras, 502, 5325

\bibitem[{{Rich} {et~al.}(2022){Rich}, {Monnier}, {Aarnio}, {Laws},
  {Setterholm}, {Wilner}, {Calvet}, {Harries}, {Miller}, {Davies}, {Adams},
  {Andrews}, {Bae}, {Espaillat}, {Greenbaum}, {Hinkley}, {Kraus}, {Hartmann},
  {Isella}, {McClure}, {Oppenheimer}, {P{\'e}rez}, \& {Zhu}}]{Rich2022}
{Rich}, E.~A., {Monnier}, J.~D., {Aarnio}, A., {et~al.} 2022, \aj, 164, 109

\bibitem[{{Rich} {et~al.}(2021){Rich}, {Teague}, {Monnier}, {Davies}, {Bosman},
  {Harries}, {Calvet}, {Adams}, {Wilner}, \& {Zhu}}]{Rich2021}
{Rich}, E.~A., {Teague}, R., {Monnier}, J.~D., {et~al.} 2021, \apj, 913, 138

\bibitem[{{Rosenfeld} {et~al.}(2014){Rosenfeld}, {Chiang}, \&
  {Andrews}}]{Rosenfeld_ea_2014}
{Rosenfeld}, K.~A., {Chiang}, E., \& {Andrews}, S.~M. 2014, \apj, 782, 62

\bibitem[{{Rosotti} {et~al.}(2020){Rosotti}, {Benisty}, {Juh{\'a}sz}, {Teague},
  {Clarke}, {Dominik}, {Dullemond}, {Klaassen}, {Matr{\`a}}, \&
  {Stolker}}]{Rosotti_ea_2020}
{Rosotti}, G.~P., {Benisty}, M., {Juh{\'a}sz}, A., {et~al.} 2020, \mnras, 491,
  1335

\bibitem[{{Sicilia-Aguilar} {et~al.}(2020){Sicilia-Aguilar}, {Manara}, {de
  Boer}, {Benisty}, {Pinilla}, \& {Bouvier}}]{Sicilia2020}
{Sicilia-Aguilar}, A., {Manara}, C.~F., {de Boer}, J., {et~al.} 2020, \aap,
  633, A37

\bibitem[{{Tang} {et~al.}(2017){Tang}, {Guilloteau}, {Dutrey}, {Muto}, {Shen},
  {Gu}, {Inutsuka}, {Momose}, {Pietu}, {Fukagawa}, {Chapillon}, {Ho}, {di
  Folco}, {Corder}, {Ohashi}, \& {Hashimoto}}]{Tang2017}
{Tang}, Y.-W., {Guilloteau}, S., {Dutrey}, A., {et~al.} 2017, \apj, 840, 32

\bibitem[{{Teague}(2019{\natexlab{a}})}]{Teague_2019_eddy}
{Teague}, R. 2019{\natexlab{a}}, The Journal of Open Source Software, 4, 1220

\bibitem[{{Teague}(2019{\natexlab{b}})}]{Teague_2019_gofish}
{Teague}, R. 2019{\natexlab{b}}, The Journal of Open Source Software, 4, 1632

\bibitem[{{Teague} {et~al.}(2022){Teague}, {Bae}, {Andrews}, {Benisty},
  {Bergin}, {Facchini}, {Huang}, {Longarini}, \& {Wilner}}]{Teague_ea_2022b}
{Teague}, R., {Bae}, J., {Andrews}, S.~M., {et~al.} 2022, \apj, 936, 163

\bibitem[{{Teague} {et~al.}(2019{\natexlab{a}}){Teague}, {Bae}, \&
  {Bergin}}]{Teague_ea_2019b}
{Teague}, R., {Bae}, J., \& {Bergin}, E.~A. 2019{\natexlab{a}}, \nat, 574, 378

\bibitem[{{Teague} {et~al.}(2018{\natexlab{a}}){Teague}, {Bae}, {Bergin},
  {Birnstiel}, \& {Foreman-Mackey}}]{Teague_ea_2018b}
{Teague}, R., {Bae}, J., {Bergin}, E.~A., {Birnstiel}, T., \& {Foreman-Mackey},
  D. 2018{\natexlab{a}}, \apjl, 860, L12

\bibitem[{{Teague} {et~al.}(2018{\natexlab{b}}){Teague}, {Bae}, {Birnstiel}, \&
  {Bergin}}]{Teague_ea_2018c}
{Teague}, R., {Bae}, J., {Birnstiel}, T., \& {Bergin}, E.~A.
  2018{\natexlab{b}}, \apj, 868, 113

\bibitem[{{Teague} {et~al.}(2019{\natexlab{b}}){Teague}, {Bae}, {Huang}, \&
  {Bergin}}]{Teague_ea_2019a}
{Teague}, R., {Bae}, J., {Huang}, J., \& {Bergin}, E.~A. 2019{\natexlab{b}},
  \apjl, 884, L56

\bibitem[{{Teague} \& {Foreman-Mackey}(2018)}]{bettermoments}
{Teague}, R. \& {Foreman-Mackey}, D. 2018, {Bettermoments: A Robust Method To
  Measure Line Centroids}, Zenodo

\bibitem[{{van der Marel} {et~al.}(2021){van der Marel}, {Birnstiel}, {Garufi},
  {Ragusa}, {Christiaens}, {Price}, {Sallum}, {Muley}, {Francis}, \&
  {Dong}}]{vanderMarel_ea_2021a}
{van der Marel}, N., {Birnstiel}, T., {Garufi}, A., {et~al.} 2021, \aj, 161, 33

\bibitem[{{van der Walt} {et~al.}(2011){van der Walt}, {Colbert}, \&
  {Varoquaux}}]{vanderWalt_np}
{van der Walt}, S., {Colbert}, S.~C., \& {Varoquaux}, G. 2011, Computing in
  Science and Engineering, 13, 22

\bibitem[{{Virtanen} {et~al.}(2020){Virtanen}, {Gommers}, {Burovski},
  {Oliphant}, {Weckesser}, {Cournapeau}, {Alexbrc}, {Peterson}, {Reddy},
  {Wilson}, {Haberland}, {Mayorov}, {Endolith}, {Nelson}, {Van Der Walt},
  {Laxalde}, {Brett}, {Polat}, {Larson}, {Millman}, {Lars}, {Van Mulbregt},
  {Eric-Jones}, {Carey}, {Moore}, {Kern}, {Leslie}, {Perktold}, {Striega}, \&
  {Feng}}]{Virtanen_scipy}
{Virtanen}, P., {Gommers}, R., {Burovski}, E., {et~al.} 2020, {scipy/scipy:
  SciPy 1.5.3}

\bibitem[{{W{\"o}lfer} {et~al.}(2022){W{\"o}lfer}, {Facchini}, {van der Marel},
  {van Dishoeck}, {Benisty}, {Bohn}, {Francis}, {Izquierdo}, \&
  {Teague}}]{Wolfer_ea_2022}
{W{\"o}lfer}, L., {Facchini}, S., {van der Marel}, N., {et~al.} 2022, arXiv
  e-prints, arXiv:2208.09494

\bibitem[{{Young} {et~al.}(2022){Young}, {Alexander}, {Rosotti}, \&
  {Pinte}}]{Young_ea_2022}
{Young}, A.~K., {Alexander}, R., {Rosotti}, G., \& {Pinte}, C. 2022, \mnras,
  513, 487

\bibitem[{{Yun} {et~al.}(2019){Yun}, {Kim}, {Bae}, \& {Han}}]{Yun_ea_2019}
{Yun}, H.-G., {Kim}, W.-T., {Bae}, J., \& {Han}, C. 2019, \apj, 884, 142

\bibitem[{{Zhang} {et~al.}(2014){Zhang}, {Isella}, {Carpenter}, \&
  {Blake}}]{Zhang_ea_2014}
{Zhang}, K., {Isella}, A., {Carpenter}, J.~M., \& {Blake}, G.~A. 2014, \apj,
  791, 42

\bibitem[{{Zhu}(2019)}]{Zhu2019}
{Zhu}, Z. 2019, \mnras, 483, 4221

\bibitem[{{Zhu} {et~al.}(2011){Zhu}, {Nelson}, {Hartmann}, {Espaillat}, \&
  {Calvet}}]{Zhu2011}
{Zhu}, Z., {Nelson}, R.~P., {Hartmann}, L., {Espaillat}, C., \& {Calvet}, N.
  2011, \apj, 729, 47

\end{thebibliography}

\begin{appendix} 
\onecolumn
\section{Observations, calibration, and imaging}
\label{app:obs}

\begin{table*}[!h]
\centering
\caption{Summary of the ALMA Band 6 observations of J1604 presented in this paper. }
\begin{tabular}{ccccccl}
\toprule
ID & EB Code & Date & Baselines & Frequency & Exp. Time & PI \\
   &     &  & [m]       &  [GHz]    & [min] &  \\
\hline
2015.1.00964.S & X412 & 2016 Jul 2  & 15-704  & 217.2-233.4 & 8.87 & Oberg\\ 
2017.A.01255.S & Xb18 & 2019 Sep 4  & 38-3638 & 213.0-230.6 & 14.49 & Benisty \\ 
               & X2fe5& 2021 Apr 29 & 15-1263 &             & 14.47 &  \\ 
               & X4583& 2019 Jul 30 & 92-8548 &             & 29.33 &  \\ 
               & X5f6e& 2019 Jul 31 & 92-8548 &             & 29.33 &  \\ 
               & X104fc& 2021 Sep 27 & 70-14362 &           & 29.33 &  \\ 
\end{tabular}
\label{tab:obsdata}
\end{table*}

To self-calibrate our observations, we proceeded as follows. We first flagged the channels containing the line to produce a continuum data set. We centered the individual execution blocks (EBs) by fitting the continuum visibilities with a ring model, allowing for a different center and amplitude.\ This enabled us to recover the phase shift and amplitude re-scaling and apply them to the EBs before combining them. To determine a good initial model for the self-calibration, we used multi-scale cleaning with the \texttt{tclean} task using a threshold of two times the rms noise level of the image. Using the tasks \texttt{gaincal} and \texttt{applycal}, we corrected for phase offsets between spectral windows, and between polarizations, considering a solution interval of the scan length (\texttt{solint=inf}). Executions obtained in 2019 were concatenated and self-calibrated together, and a similar procedure was applied to those obtained in 2021. In addition to the first round of self-calibration, two additional iterations of phase self-calibration were done, with solution intervals of 300s and 180s for the 2019 data; only one round was done for the 2021 data, with a solution interval of 360s. For both data sets, a round of amplitude self-calibration was applied with \texttt{solint=inf}. The solutions were then applied to the gas data. While these two epochs will be analyzed separately for the continuum in a forthcoming paper (Kurtovic et al.), to analyze the gas data, we concatenated them after checking that the data do not show significant variations between the two epochs. We imaged the resulting visibilities with the \texttt{tclean} task using the multi-scale \texttt{CLEAN} algorithm with scales of 0, 1, 3, and 6 times the beam FWHM and an elliptic \texttt{CLEAN} mask encompassing the disk emission. The $^{12}$CO (2-1) molecular line observations are imaged with a robust value of 1.0 and a channel width of 0.1\,\kmsec and masked by the 4.0\,$\sigma$ threshold. The data were tapered to 0.\arcsec{}1, and we used the ``JvM correction'' \citep{jvm1995, czekala2021}.


\begin{figure*}[h]
\centering
\includegraphics[width=0.95\linewidth]{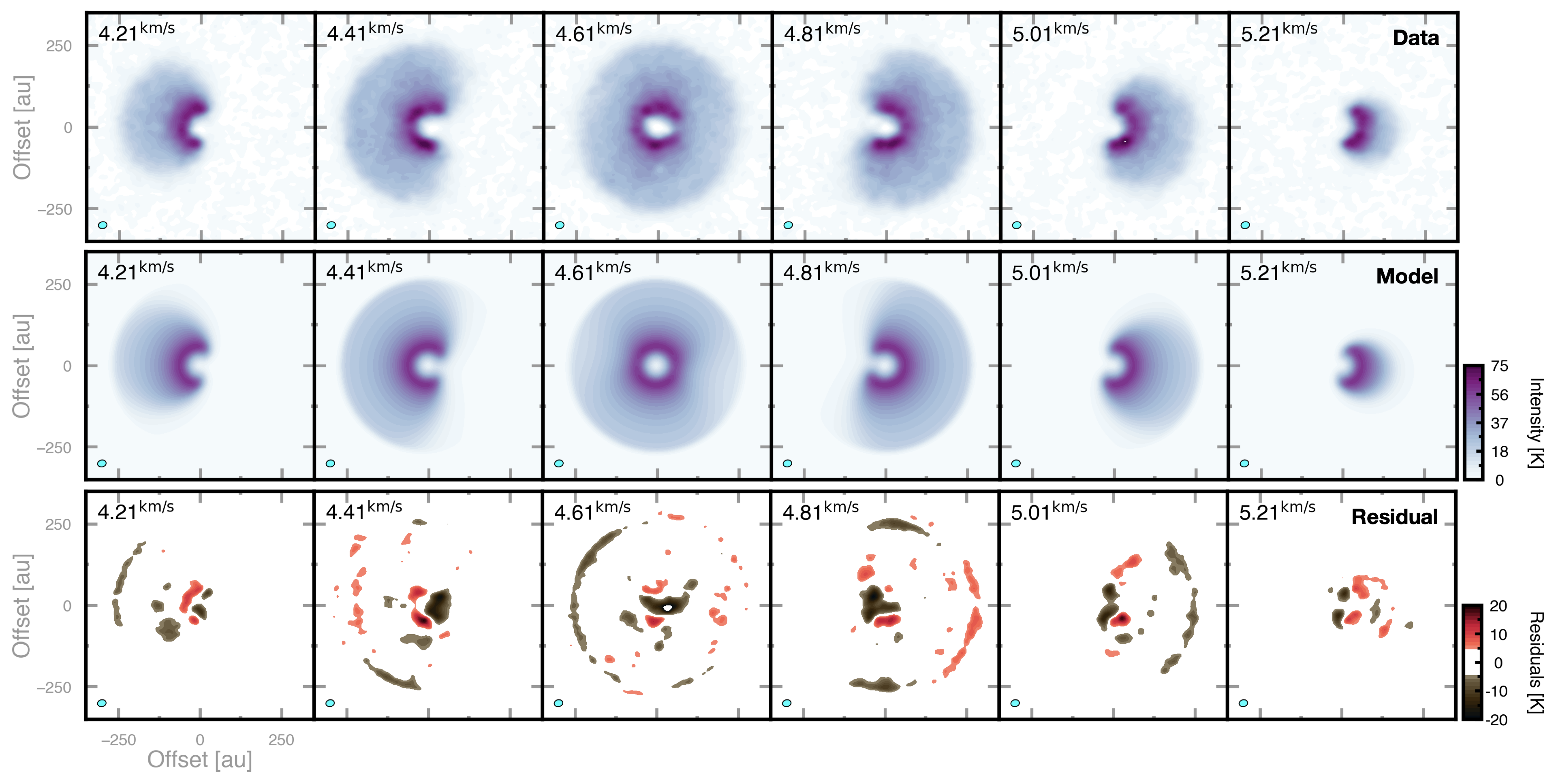}
\caption{Gallery of selected channel maps. Panels show the $^{12}$CO data channel maps (top row) and best-fit model channel maps (middle row), together with intensity residuals in kelvins for each channel (bottom row). In the bottom row the colorbar has been adjusted such that residuals smaller than 1$\sigma$ are white. The beam size is depicted in the lower-left corner of each channel. For reference, the best-fit systemic velocity was found to be $v_\mathrm{sys}=4.62$\,\kmsec and the channel spacing is 100\,\msec.}
\label{fig:app_channel_maps_compared}
\end{figure*}

\begin{figure}[h!]
\centering
\includegraphics[width=0.6\linewidth]{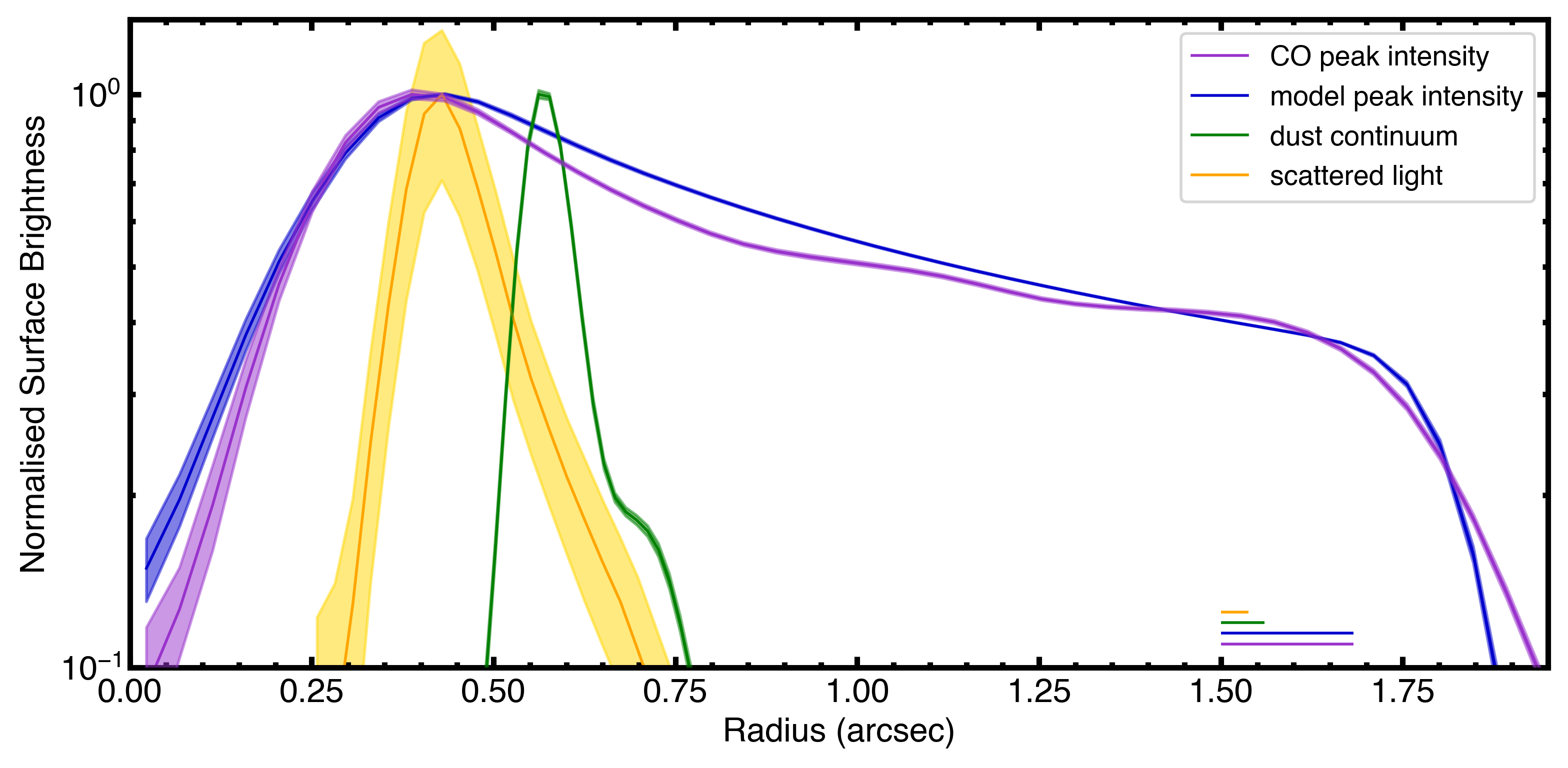}
\caption{Radial profile of the surface brightness for different tracers. Profiles are normalized to the peak of the emission for the 231 GHz continuum and the CO peak flux (both for the data and \texttt{discminer} model) and for the SPHERE scattered-light observation. Shaded regions show the standard deviation of each azimuthal average. The lines in the lower-right corner show the major beam size (resolution) for each profile in the corresponding color.}
\label{fig:app_I_azim_avg}
\end{figure}

\begin{figure}[h!]
\centering
\includegraphics[width=0.6\linewidth]{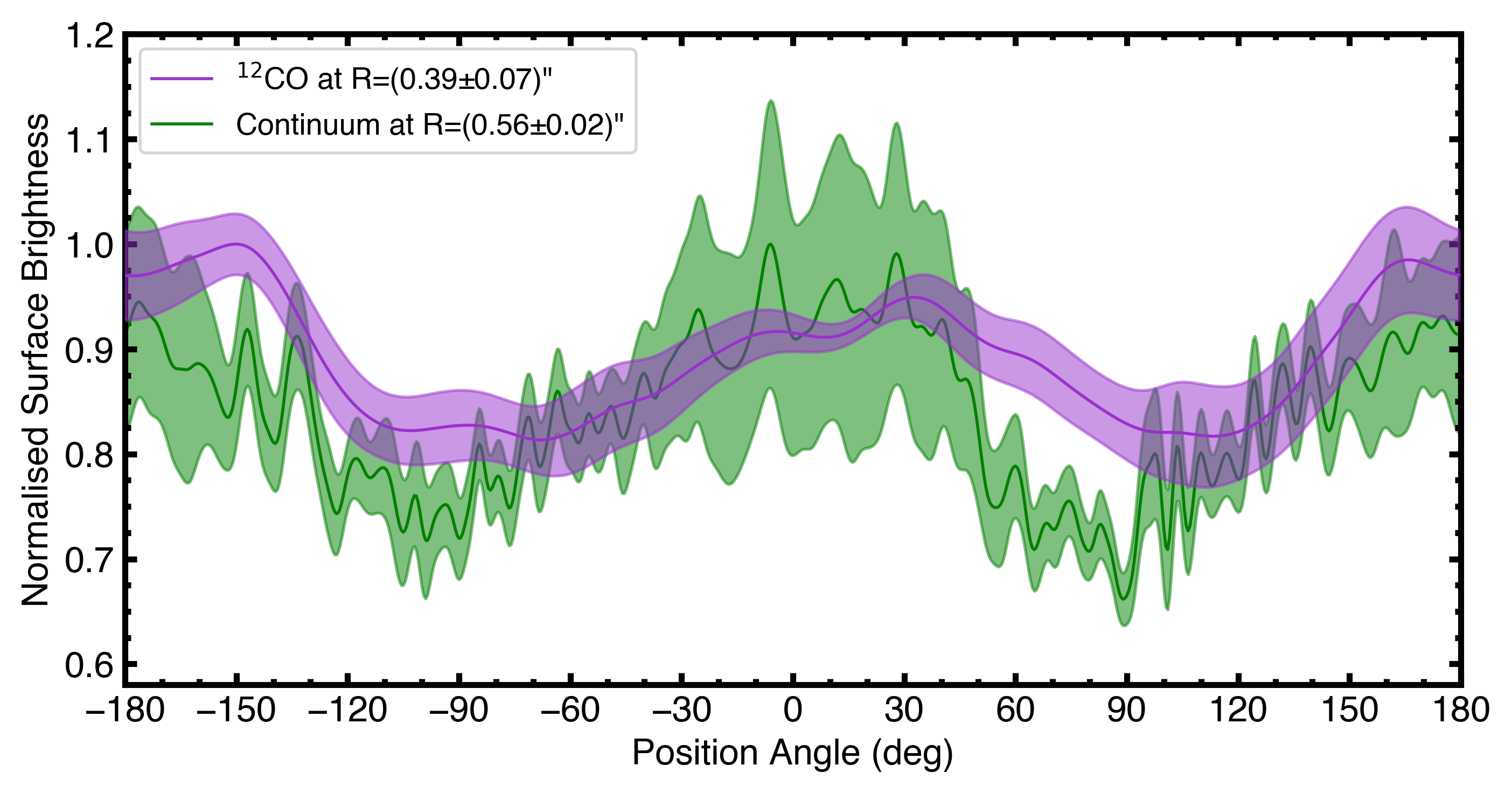}
\caption{Azimuthal profiles of the surface brightness, normalized to the peak of the emission. Profiles were extracted at an annulus with a width of approximately one corresponding beam size centered at 0.56\arcsec{} and 0.39\arcsec{} for the 231 GHz continuum and the CO peak flux, which both show significant azimuthal intensity variations, of 34\% and 19\%, respectively. Shaded regions show the standard deviation of each radial average.}
\label{fig:app_azim_int}
\end{figure}

\begin{figure*}[h!]
\centering
\includegraphics[width=0.95\linewidth]{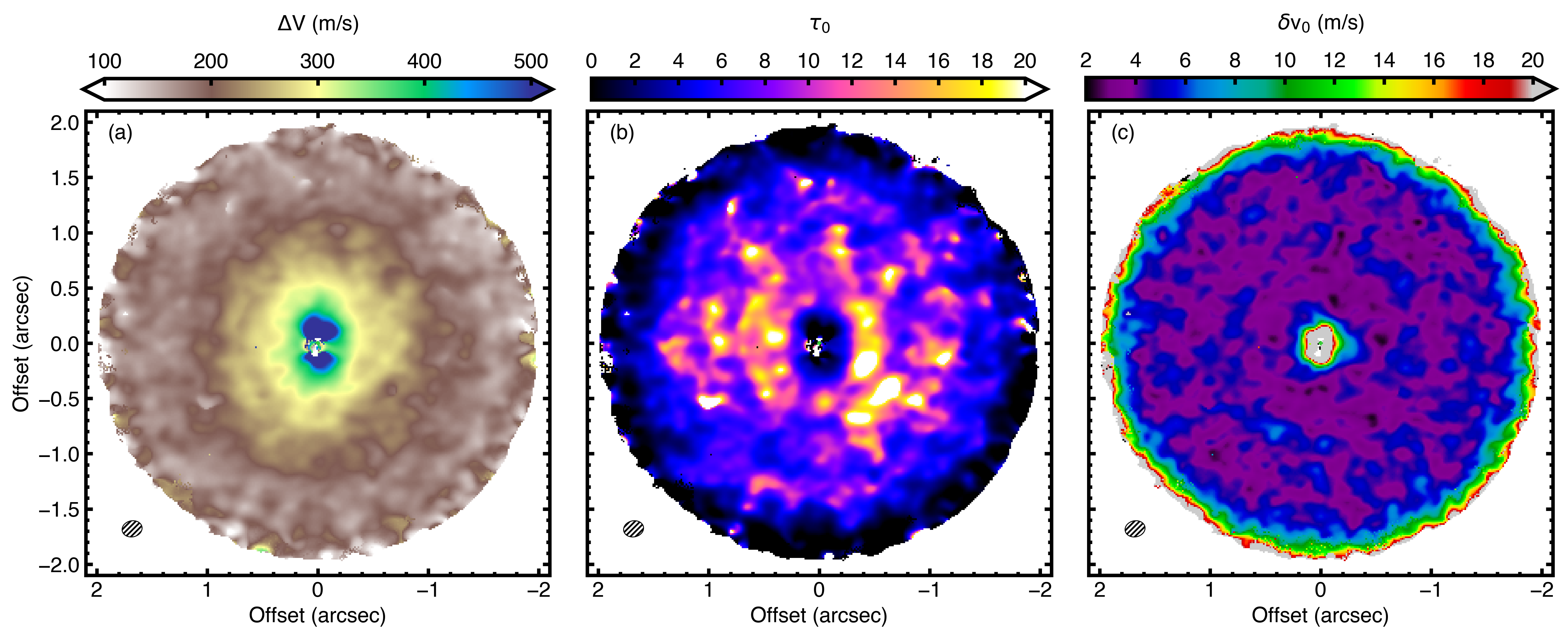}
\caption{Additional moment maps of the centroid fitting. Panels show the line width, $\Delta V$ (left), the peak optical depth, $\tau_0$ (center), and the error of the centroid fitting, $\delta v_0$ (right). Note that for $\tau_0<1$ one can assume the line profile to be well presented by a Gaussian, while for $\tau_0>5$ the line profile has a saturated core, i.e., a very flat top (see Eq.\,\ref{eq:Gauss_thick}). The beam size is depicted in the lower-left corner, and only regions where $I_0>5\sigma$ with $\sigma=1.1$\,mJy\,beam$^{-1}$ are shown.}
\label{fig:app_moment_maps}
\end{figure*}

 \begin{figure}[h!]
\centering
\includegraphics[width=0.6\linewidth]{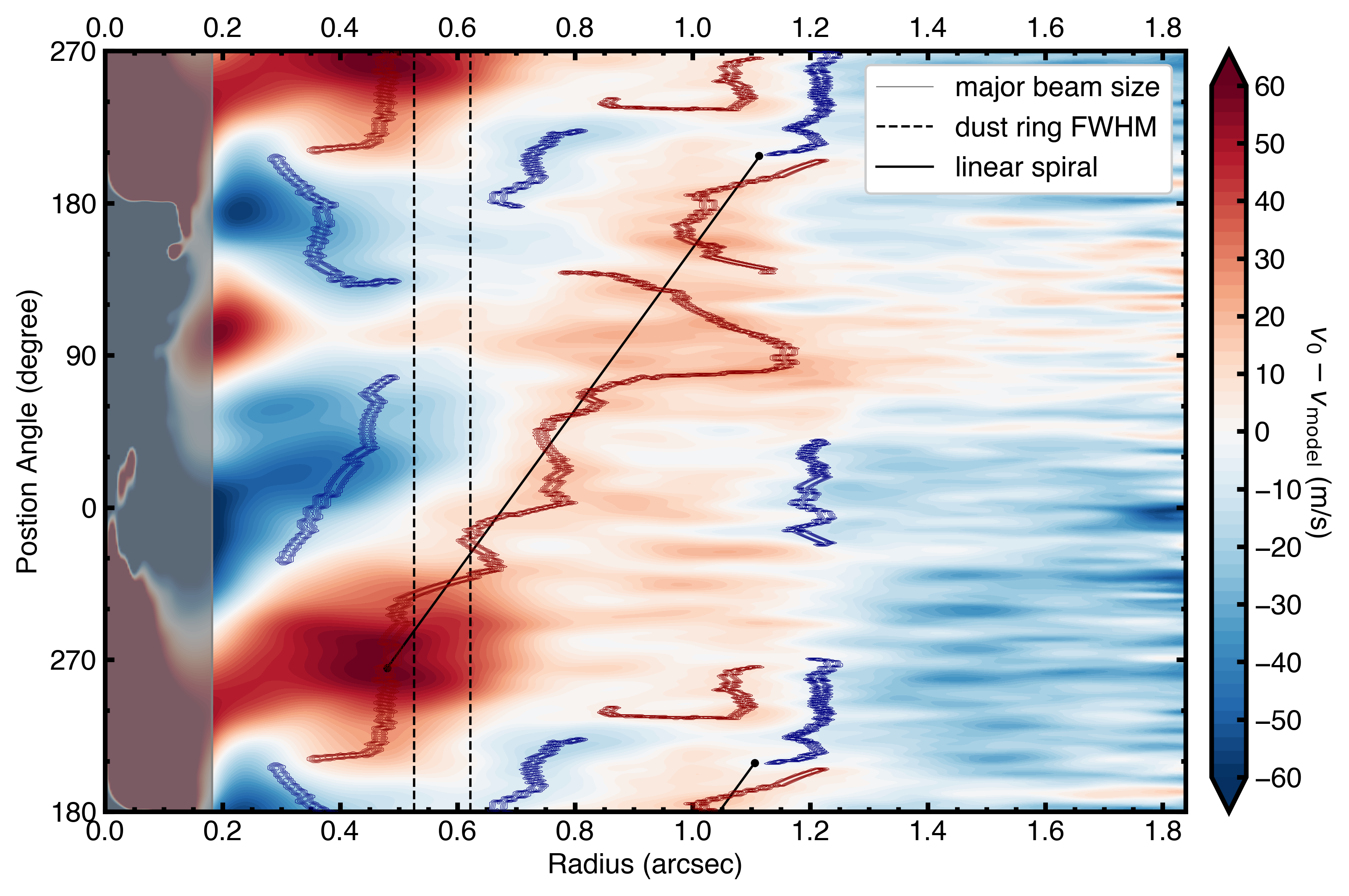}
\caption{Polar map of the velocity residuals. Same as Fig. \ref{fig:polar_map}, but now overlaid with filamentary structures found by \texttt{FilFinder}. The overplotted red and blue lines are the medial axes of the filamentary structures found by the algorithm. To trace the apparent spiral in the residuals, we restricted the algorithm to search for filaments in the radial locations $r=[0.3, 1.25]$\arcsec{}. For the filamentary detection, we assumed a smoothing size of one synthesized beam size and a minimum size of 500 pixels for a filament to be considered.}
\label{fig:app_polar_filam_resid}
\end{figure}

\begin{figure}[h!]
\centering
\includegraphics[width=0.65\linewidth]{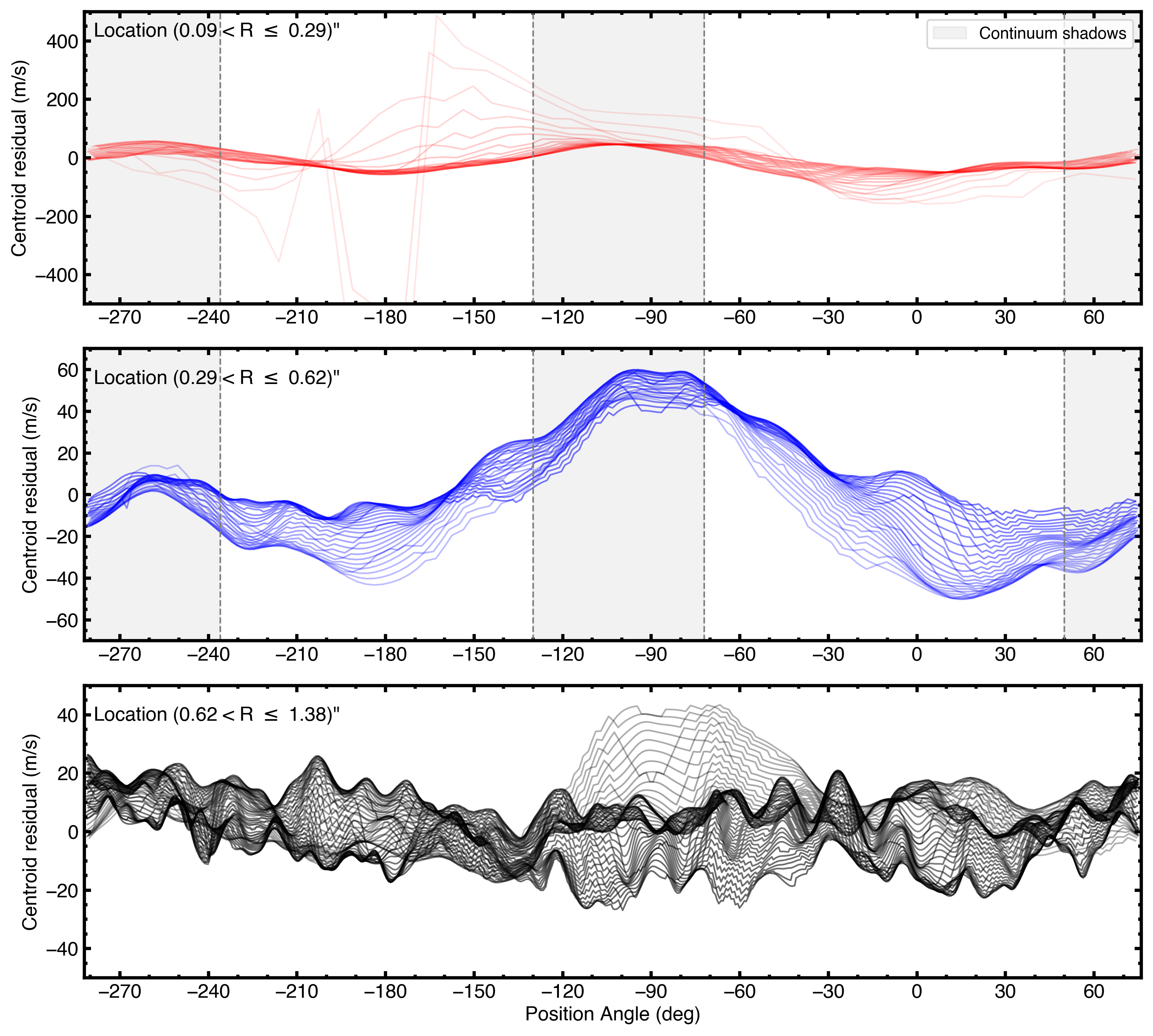}
\caption{Polar contour map of the centroid residuals. The upper panel shows residuals inside the cavity, the middle panel residuals between the cavity and the outer edge of dust ring, and the lower panel the residuals in the outer disk. The radial spacing between each contour is $\sim$1.8\,au, and the opacity of the lines increases with radius. We would like to emphasize that the bump between PA$\approx-(110-70)^\circ$ in the middle panel includes most of the residual points we detect with the peak variance method of \texttt{discminer}, which can be readily seen in the left panel of Fig.~\ref{fig:app_cluster_r_phi}.}
\label{fig:app_contour_resid}
\end{figure}

\begin{figure}[h!]
\centering
\includegraphics[width=0.95\linewidth]{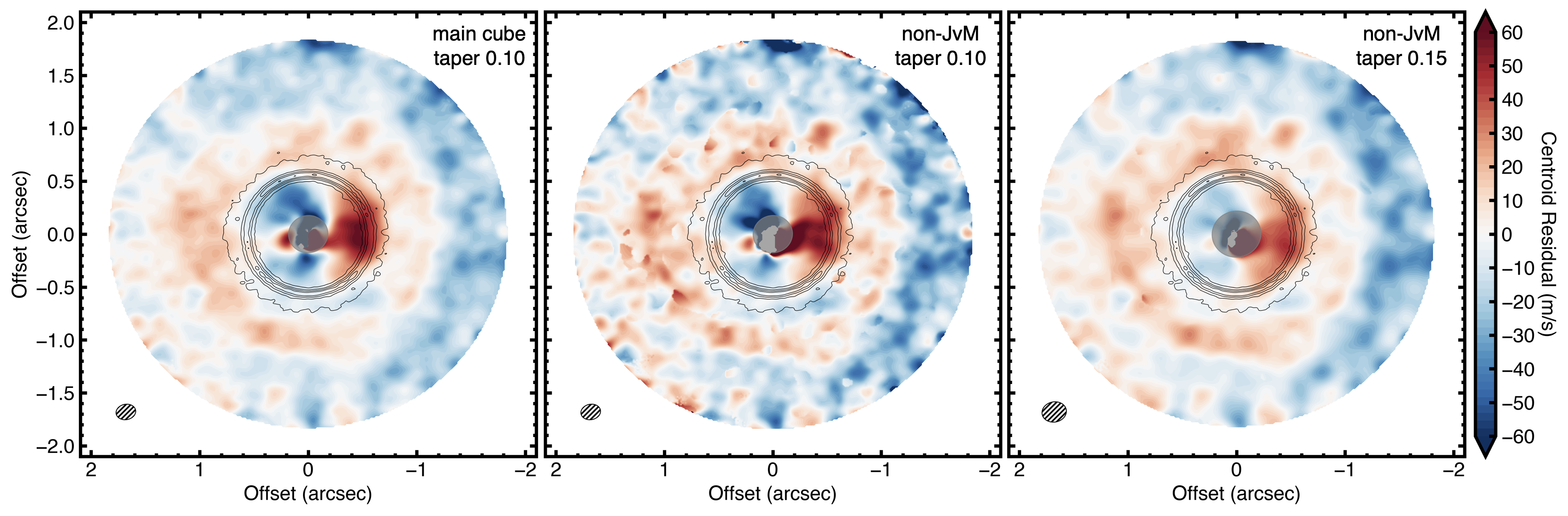}
\caption{ Comparison of the velocity residual maps for different imaging parameters. The left panel corresponds to the cube used in the main text, while the middle panel corresponds to the same cube without JvM correction. The right panel shows the residual maps for a non-JvM-corrected cube with a different taper (0.15\arcsec{}). In all panels the dust continuum is overlaid in solid contours with equal levels, as in Fig. \ref{fig:intensities}. Best-fit Keplerian models were subtracted from each of the cubes.  The detection of the non-Keplerian features is quite robust, irrespective of the imaging procedure. }
\label{fig:app_nonJvM_residuals}
\end{figure}
\newpage

In Fig.\,\ref{fig:app_nonJvM_residuals} we show a comparison of velocity residual maps for additional cubes, imaged using different imaging parameters, to assess the robustness of our detections. We compare residual maps for the same cube as used in the main text,  but without JvM correction, and for a cube imaged with a different tapering (0.15\arcsec{} instead of 0.10\arcsec{}). Best-fit Keplerian models were subtracted from each of the cubes. As is evident from the comparison of the residual maps, the detection of the reported non-Keplerian features are robust irrespective of the imaging parameters. However, the detailed morphology of the velocity residual peaks changes with imaging parameters and, as a consequence, the value of the inferred planet location from the discminer analysis. We find that the best-fit discminer models to the non-JvM-corrected cubes are similar within 3\% with respect to the best-fit parameters listed in Appendix \ref{app:mcmc_model}, with the exception of the line slope and some of the peak intensity parameters, which vary by up to 15\%. While estimating the systematics due to imaging parameters is beyond the scope of this Letter, Fig.\,\ref{fig:app_nonJvM_residuals} provides evidence that the detection of non-Keplerian features is robust. We measure the rms in a line-free channel to be 2.6\,mJy\,beam$^{-1}$ and 2.9\,mJy\,beam$^{-1}$ for the non-JvM-corrected cubes with a taper of 0.10 and 0.15, respectively.

\section{Model best-fit parameters}
\label{app:mcmc_model}

    \begin{table*}[h!]
    \centering
    \begin{threeparttable}
    \renewcommand{\arraystretch}{2}
    \begin{tabular}{l l l l l l}
    \toprule
    Attribute & \multicolumn{2}{c}{Prescription} & \multicolumn{2}{c}{Best-fit parameters} \\
    \hline
    Centre offset & $x_\mathrm{c}$, $y_\mathrm{c}$&  & $x_\mathrm{c}=-2.66^{+0.05}_{-0.06}$\,au & $y_\mathrm{c}=-0.07\pm0.03$\,au \\
    Position angle & PA & & PA\,$=258.75^{+0.06}_{-0.05}$ deg & -  \\
    \hline
   Systemic velocity & $v_\mathrm{sys}$ & & $v_\mathrm{sys}=4617.2^{+0.3}_{-0.4}$\,\msec & -  \\
   Rotation velocity & $v_\mathrm{kep}=\sqrt{\frac{GM_\star}{R}}$ & & $M_\star=1.220\pm0.001\,M_\odot$ & -  \\
   \hline
    & $I_{p}=I_{p0}\,(R/R_\mathrm{break})^{p_0} $ & $R\leq R_\mathrm{break}$ & $I_{p0}=9.388^{+0.003}_{-0.005}$\,mJy\,pixel$^{-1}$ &$p_0=1.497^{+0.004}_{-0.005}$ \\
   Peak intensity & $I_{p}=I_{p0}\,(R/R_\mathrm{break})^{p_1} $ & $R_\mathrm{break} < R \leq R_\mathrm{out} $ & $R_\mathrm{break}=56.78^{+0.06}_{-0.05}$\,au & $p_1=-0.789\pm0.001$  \\
    & $I_{p}=0$ & $R < R_\mathrm{out} $ &  $R_\mathrm{out}=267.2\pm0.1$\,au & - \\
   Line width & $L_w=L_{w0}(R/D_0)^p$ & & $L_{w0}=0.4097\pm0.0004$\,\kmsec & $p=-0.592^{+0.001}_{-0.002}$ \\
   Line slope & $L_s=L_{s0}(R/D_0)^p$ & & $L_{s0}=4.569^{+0.008}_{-0.009}$ & $p=-0.454^{+0.005}_{-0.008}$ \\
    \hline
    \end{tabular}
    \end{threeparttable}
    \renewcommand{\arraystretch}{1}
    \caption[]{Attributes of the \texttt{discminer} model for the $^{12}$CO intensity channel maps of the disk around J1604. PA is the position angle of the semimajor axis of the disk on the redshifted side, R the cylindrical radius, and $D_0 = 100$\,au a normalization constant for the line properties. The (down-sampled) pixel size of the model is 8.8\,au.}
    \label{table:best_fit_param}
    \end{table*}
For the initial \texttt{emcee} run, we used literature values for the position angle and stellar mass \citep[PA=260$^\circ$, $M_\star=1.24\,M_\odot$;][respectively]{Dong2017,Manara_ea_2020}. The initial values of the other parameters were found by comparing the overall morphology between the data and a prototype model. We performed the MCMC search with 150 walkers that evolved for 2000 steps for an initial burn-in stage. We proceeded in two steps. First, we masked the disk region inward of the dust continuum and only fitted the outer disk ($R>90\,$au) to get a robust estimate of the stellar mass and avoid the code being confusing by the strongly non-Keplerian velocity features in the inner regions. In this run, we interestingly find a strong offset from the disk center in the x direction of $-8.0\,$au. In a second step, we fixed the stellar mass and now fitted for the whole disk, masking an inner region corresponding to one major beam size (26\,au) in radius, where effects of beam smearing are strongest. We ran 150 walkers for 
20000 steps until reaching convergence, which presented as a nearly normal distribution of the walkers. The variance and median of the parameter walkers remain effectively unchanged after $\sim7000$ steps. The best-fit parameters are the median of the posterior distributions, and given errors are the 16 and 84 percentiles in the last 5000 steps of the 20000-step run, summarized in Table~\ref{table:best_fit_param}.

\begin{figure}[h!]
\centering
\includegraphics[width=0.95\linewidth]{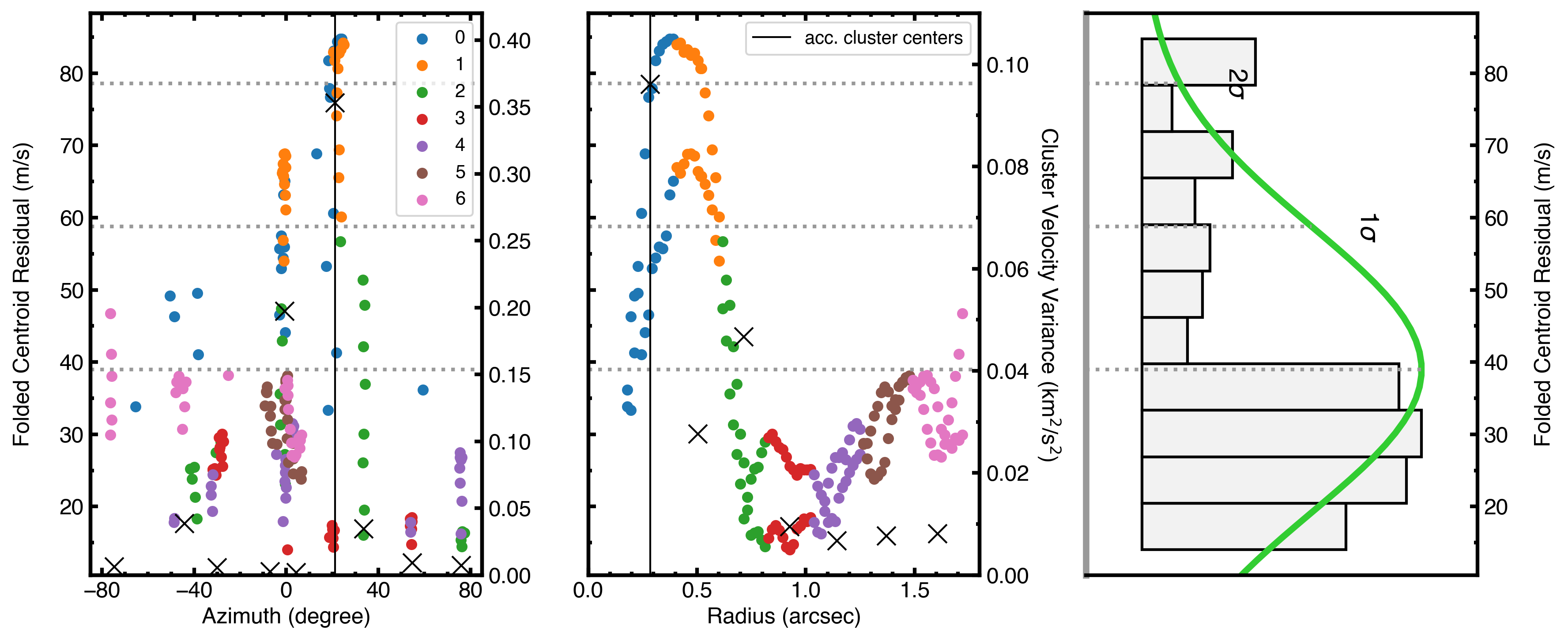}
\caption{Location of the folded peak velocity residuals. The detected points are shown in azimuth (left) and radius (middle) and were obtained using the peak variance method. Colors indicate the seven different radial clusters specified, where blue peak residual points are within a detected significant radial cluster. The black crosses are the velocity variances of the clusters plotted at the ($R,\phi$) location of each cluster center. The centers of the accepted clusters (those with peak velocity residuals larger than three times the variance in other clusters) in radius and azimuth are marked with vertical black lines in both panels. The right-hand plot shows the normal distribution of the peak residual points in a histogram. Note that outliers of the distribution are related to the localized perturbation. The maximum value of all peak folded centroid residuals is at 0.39\arcsec{} (57\,au), its mean value is 39\,m/s, and $1\sigma_v=20$\,m/s (not to be mistaken with the cluster variances).}
\label{fig:app_cluster_r_phi}
\end{figure}

\section{Decomposition and deprojection of velocity components}
\label{app:vel_depr}
\begin{figure}[h!]
\centering
\includegraphics[width=0.6\linewidth]{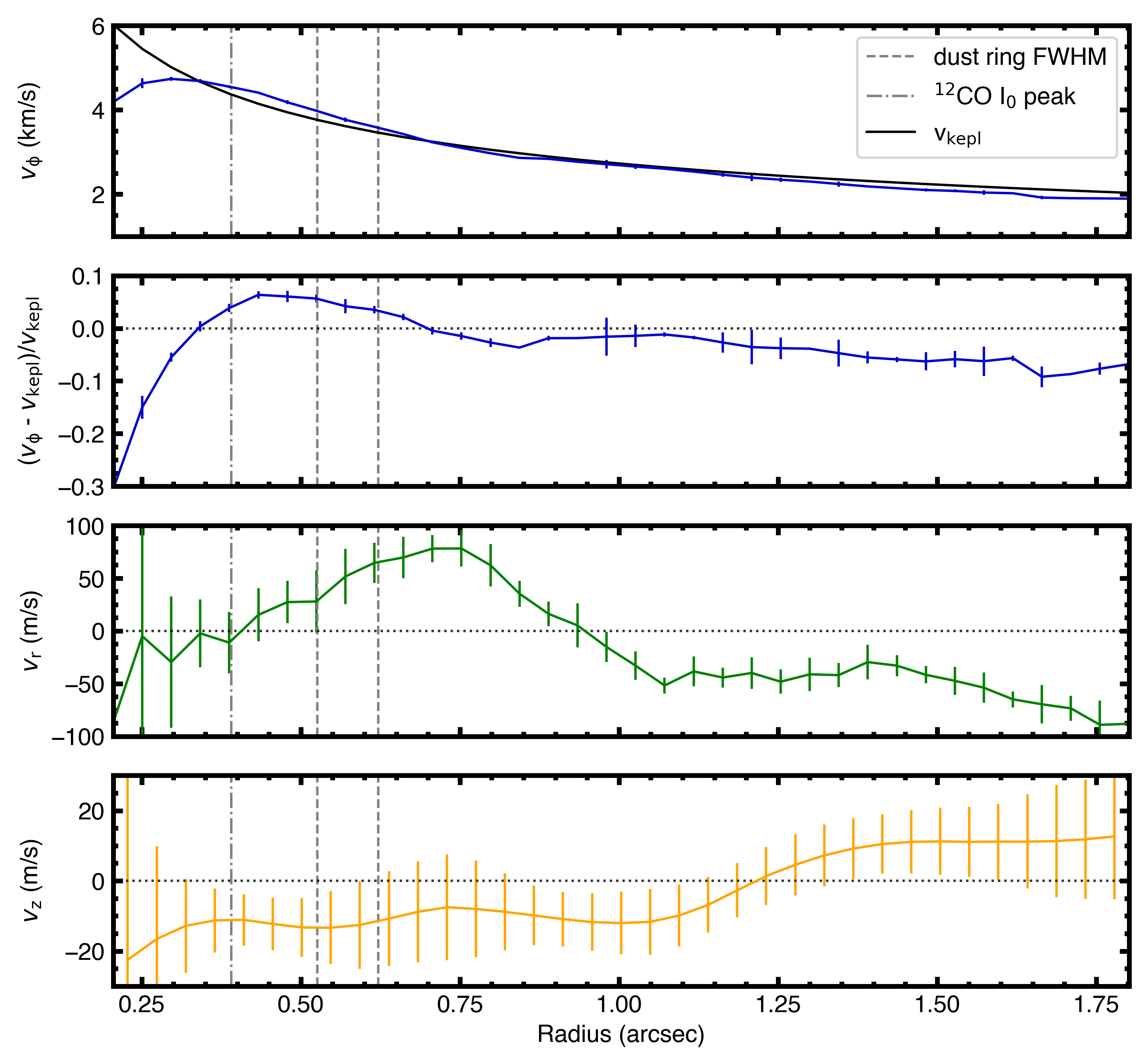}
\caption{Azimuthally averaged and deprojected azimuthal, radial, and vertical velocity components. The radial width of each annulus is one-fourth of the synthesized beam size. The error bars are given by the standard deviation for each velocity component averaged over the 20 iterations used.}
\label{fig:app_depr_velocities}
\end{figure}

To determine the rotation curves of each velocity component, we used the code \texttt{eddy} \citep{Teague_2019_eddy}. We followed the method presented in \cite{Teague_ea_2018c}, which uses a Gaussian process to determine the azimuthal, $v_\phi$, and radial, $v_r$, velocity components along a given annulus. 
To this end, we divided the disk into concentric annuli with a radial width of one-fourth of the synthesized beam ($\sim0.$05\arcsec{}), which ranges from 0.18\arcsec{} to 1.85\arcsec{}, and extracted the velocities over 20 iterations to minimize their standard deviation. To obtain the vertical velocity component, $v_z$, we used the measured azimuthally averaged profiles of $v_\phi$ and $v_r$ and extended them to produce 2D maps, considering the projection of the following components along the line of sight:
\begin{equation}
\label{eq:v_deproj}
\begin{aligned}
    v_{\phi,\,\mathrm{proj}} &= v_\phi\,\cos{(\phi)}\,\sin{(|i|)}, \\
    v_{r,\,\mathrm{proj}} &= v_r\,\sin{(\phi)}\,\sin{(i)}, \\
    v_{z,\,\mathrm{proj}} &= -v_z\,\cos{(i)},
\end{aligned}
\end{equation}
where $\phi$ is the polar angle in the disk frame (such that $\phi$= 0 corresponds to the redshifted major axis) and $i$ the inclination of the disk. In the case of J1604, the disk rotates clockwise, which corresponds to a negative inclination in the above definition. We then subtracted these maps together with the systemic velocity, $v_\mathrm{sys}$, from the line-of-sight velocity, $v_0$, map (Fig.~\ref{fig:centroid},a) to obtain a map of the vertical velocity component, $v_{z,\,\mathrm{proj}}$:
\begin{equation}
\label{eq:v_z_sky}
    v_{z,\,\mathrm{proj}}=v_0-v_\mathrm{sys}-v_{\phi,\,\mathrm{proj}}-v_{r,\,\mathrm{proj}}.
\end{equation}
The radial profile of $v_z$ was obtained by deprojecting and azimuthally averaging its 2D velocity map. The radial profiles of the deprojected velocity components can be found in Fig.~\ref{fig:app_depr_velocities}.

\begin{figure*}[h!]
\centering
\includegraphics[width=0.85\linewidth]{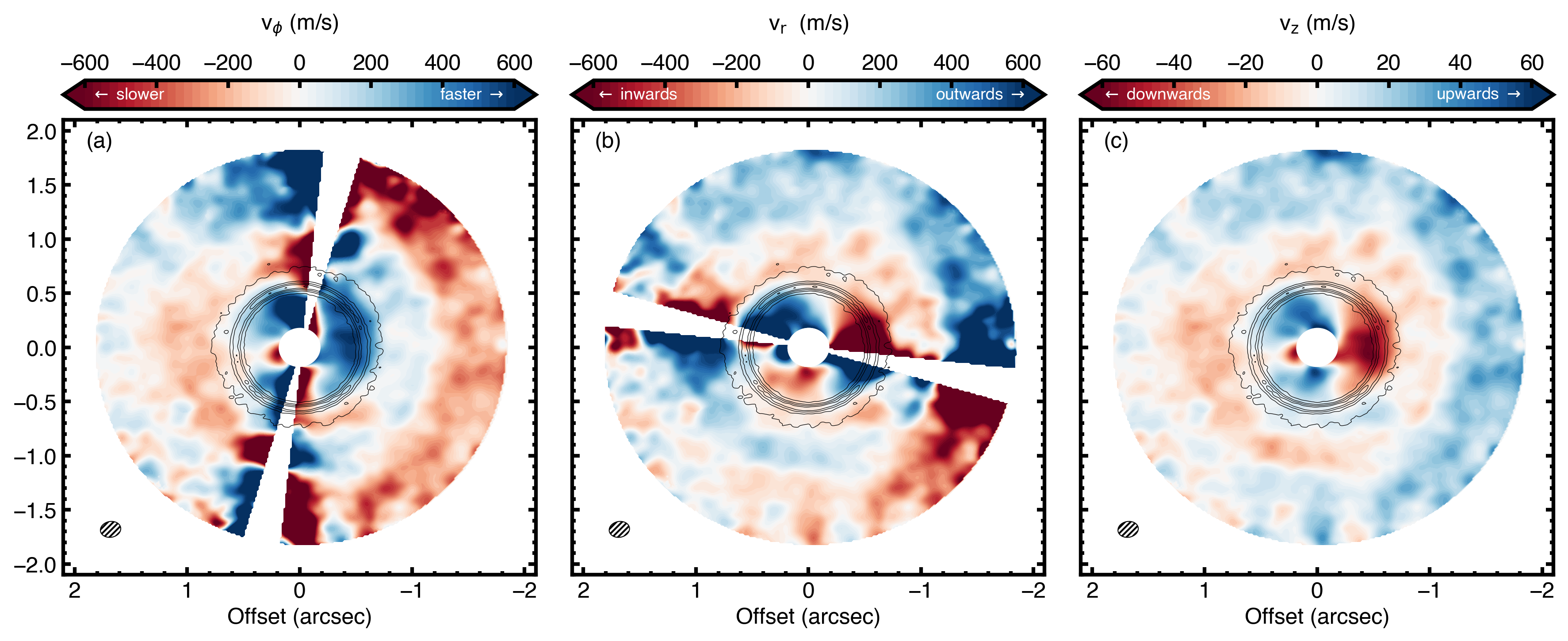}
\caption{J1604 deprojected velocity components. It is assumed that all velocities are azimuthal (left column), radial (central column), or vertical (right column). For the azimuthal and radial components, wedges along the minor and major axis have been masked as the observations are insensitive to these components (see Eq.~\ref{eq:v_deproj}). In each panel the synthesized beam is shown in the lower-left corner.}
\label{fig:app_2D_depr_velocities}
\end{figure*}

\clearpage
\section{Derivation of $\Delta v_\phi$ as a function of azimuthal temperature variations, $\Delta$T}
\label{app:nav_st_deriv}
To relate the change in the brightness temperature, $\Delta T$, of $^{12}$CO to variations in the rotational velocity, $v_\phi$, we solved the Navier-Stokes equation in cylindrical coordinates in the $\phi$ direction:
\begin{equation}
\begin{aligned}
\rho_g \frac{v_\phi}{R} \frac{\partial v_\phi}{\partial \phi} =& \frac{1}{R} \frac{\partial p}{\partial \phi} + \mu \left(\frac{\partial}{\partial R}\left(\frac{1}{R}\frac{\partial}{\partial R}(R v_\phi)\right)\frac{1}{R^2}\frac{\partial^2 v_\phi}{\partial \phi^2}\right),
\end{aligned}
\end{equation}
where  $R$ is the cylindrical radius, $\rho_g$ the gas density, and  $\mu$ the mean molecular weight. In a first step, we assumed that the radial variations within the chosen annulus are negligible and inserted the disk gas pressure in the vertically isothermal assumption: $p=\rho_g c_s^2 = \rho_g \frac{k_\mathrm{B} T}{\mu m_p}$, where $k_\mathrm{B}$ is the Boltzmann constant and  $m_p$ the proton mass. In a second step, we further assumed the gas density to be constant along the annulus $\rho_g=\mathrm{const.}$ and re-arranged the equation:
\begin{equation}
\begin{aligned}
\rho_g \frac{v_\phi}{R} \frac{\partial v_\phi}{\partial \phi} =& \frac{1}{R} \frac{\partial}{\partial \phi} \left(\rho_g \frac{k_\mathrm{B} T}{\mu m_p}\right) +  \frac{\mu}{R^2}\frac{\partial^2 v_\phi}{\partial \phi^2} \\
\frac{\partial v_\phi}{\partial \phi} - \frac{\mu}{v_\phi R \rho_g}\frac{\partial^2 v_\phi}{\partial \phi^2} =& \frac{k_\mathrm{B}}{v_\phi \mu m_p} \frac{\partial T}{\partial \phi} \\
\frac{\partial v_\phi}{\partial \phi} - \frac{0.4 \mu}{v_\phi \sqrt{2\pi}\Sigma_g}\frac{\partial^2 v_\phi}{\partial \phi^2} =& \frac{k_\mathrm{B}}{v_\phi \mu m_p} \frac{\partial T}{\partial \phi}
\end{aligned}
.\end{equation}
In the last step, we inserted the gas midplane density, $\rho_g=\Sigma_g/(\sqrt{2\pi}H)=\Sigma_g/(\sqrt{2\pi}\,0.2 R),$ assuming a disk aspect ratio of $H/R=0.2$. Assuming that $v_\phi \approx v_\mathrm{kep}$ and $\Sigma_g \approx 1$g/cm$^2$ \citep[see Fig.\,3 of][]{Dong2017} at the location of the annulus at $R\sim0.4$\arcsec{} (58\,au), we can assess the order of magnitude of the second term on the left-hand side of the equation, which is only on the order of $10^{-6}$. Therefore, we neglected the second-order derivative and identified the sound speed, $c_s$:
\begin{equation}
\begin{aligned}
\Delta v_\phi \approx & \frac{k_\mathrm{B}}{v_\mathrm{kep}\, \mu m_p} \frac{T}{T} \Delta T_\phi \approx \frac{c_s^2}{v_\mathrm{kep}}\frac{\Delta T_\phi}{T} \,\Big \rvert \div v_\mathrm{kep}\\
\frac{\Delta v_\phi}{v_\mathrm{kep}} \approx & \left(\frac{c_s}{v_\mathrm{kep}}\right)^2 \frac{\Delta T_\phi}{T} \approx \left(\frac{H}{R}\right)^2\frac{\Delta T_\phi}{T}
\end{aligned}
.\end{equation}
Equation D.3 now connects the fractional azimuthal temperature variation, $\Delta T_\phi/T$, to the rotational velocity deviation relative to Keplerian, $\Delta v_\phi/v_\mathrm{kep}$.



\end{appendix}

\end{document}